\documentclass[utf8,latin1,aps,showpacs,twocolumn,floats,superscriptaddress,longbibliography]{revtex4-1}

\usepackage{dcolumn}

\usepackage[pdftex]{graphicx}

\usepackage[UKenglish]{babel}
\usepackage{amsmath,amssymb,euscript}
\usepackage{mathtools}
\usepackage{placeins}

\RequirePackage[
hyperindex,colorlinks,bookmarksnumbered,
plainpages=true,pdfstartview=FitH]{hyperref}
\hypersetup{linkcolor=blue,urlcolor=blue,citecolor=blue}
\usepackage{hyperref}

\usepackage{braket}
\usepackage{bbm}
\usepackage{mathrsfs}

\usepackage{ulem}

\newcommand{\pdag}{{\phantom{\dagger}}}

\def\ii{i}
\def\e{e}
\def\up{\uparrow}
\def\down{\downarrow}
\def\thetarot{\theta_q}

\def\Im{\text{Im}}
\newcommand\ds[1]{\displaystyle{#1}}

\usepackage{color}
\definecolor{darkgreen}{rgb}{0,0.5,0}
\definecolor{darkred}{rgb}{0.5,0,0}

\newcommand{\Eq}[1]{Eq.~\eqref{#1}}
\newcommand{\Fig}[1]{Fig.~\ref{#1}}

\usepackage{hyperref}

\begin{document}
\title{Nonequilibrium Steady-State Transport in Quantum Impurity Models: \\
a Thermofield and Quantum Quench Approach  using
Matrix Product States}
\author{F.~Schwarz}\affiliation{Physics  Department,  Arnold  Sommerfeld  Center  for  Theoretical  Physics,  and  Center  for  NanoScience, Ludwig-Maximilians-Universit\"at,  Theresienstra{\ss}e  37,  80333  M\"unchen,  Germany}
\author{I.~Weymann}\affiliation{Faculty of Physics, Adam Mickiewicz University, Umultowska 85, 61-614 Pozna\'n, Poland}
\author{J.~von Delft}\affiliation{Physics  Department,  Arnold  Sommerfeld  Center  for  Theoretical  Physics,  and  Center  for  NanoScience, Ludwig-Maximilians-Universit\"at,  Theresienstra{\ss}e  37,  80333  M\"unchen,  Germany}
\author{A.~Weichselbaum}\affiliation{Physics  Department,  Arnold  Sommerfeld  Center  for  Theoretical  Physics,  and  Center  for  NanoScience, Ludwig-Maximilians-Universit\"at,  Theresienstra{\ss}e  37,  80333  M\"unchen,  Germany}

\begin{abstract}
  The numerical renormalization group (NRG) is tailored to describe
  interacting impurity models in equilibrium, but faces limitations
  for steady-state nonequilibrium, arising, e.g., due to an applied
  bias voltage.  We show that these limitations can be overcome by
  describing the thermal leads using a thermofield approach,
  integrating out high energy modes using NRG, and then treating the
  nonequilibrium dynamics at low energies using a quench protocol,
  implemented using the time-dependent density matrix renormalization
  group (tDMRG).  This yields quantitatively reliable results
  for the current (with errors $\lesssim 3\%$) down to the exponentially small energy scales characteristic of
  impurity models.  We present results of benchmark quality
  for the temperature and
  magnetic field dependence of the zero-bias conductance peak for the
  single-impurity Anderson model.
\end{abstract}

\maketitle

\textit{Introduction.---}
A major open problem in the theoretical study of nanostructures such
as quantum dots or nanowires is the reliable computation of the
nonlinear conductance under conditions of nonequilibrium
steady-state (NESS) transport. These are open quantum systems
featuring strong local interactions, typically described by quantum
impurity models such as the interacting resonant level model (IRLM),
the Kondo model (KM) or the single-impurity Anderson model (SIAM).
Much work has been devoted to studying the NESS properties of such
models using a variety of methods
\cite{Rosch03,Kehrein05,Anders08,Kirino08,Meisner09,Eckel10,Werner10,Pletyukhov12,Smirnov13,Cohen14,Antipov16,Reininghaus14,Dorda15,Jakobs07,Boulat08},  leading to a fairly good qualitative understanding of their behavior. The interplay of strong correlations, NESS driving and dissipative effects leads to a rich and complex phenomenology. In particular, for the KM and SIAM, the nonlinear conductance 
exhibits  a striking zero-bias peak, the so-called Kondo peak, characterized by a
 small energy scale, the Kondo temperature $T_K$, that weakens with increasing temperature and splits with increasing magnetic field, in qualitative agreement with experiments \cite{Ralph94,Goldhaber98,Cronenwett98,Simmel99,vanderWiel00,Kretinin11,Kretinin12}.  However, a full, quantitative description of the NESS behavior of such models under generic conditions has so far been unfeasible: none of the currently available approaches
 meet the threefold challenge of (i) treating interactions essentially exactly, (ii) resolving very small energy scales, 
and (iii) incorporating NESS conditions.

This Letter presents an approach that does meet
this
challenge. (i) To deal with interactions, we use numerical matrix
product state (MPS) methods. (ii) We use the numerical renormalization group
(NRG) \cite{Wilson75,Bulla08} to 
integrate out high-energy
modes, leading to a renormalized impurity problem \cite{Guttge2013}
whose reduced effective bandwidth, $D^\ast$, is set by a transport
window defined by the voltage bias 
($V$) and the temperature
($T$). This considerably
enlarges the window of 
accessible time scales, which scale as $1/D^\ast$,
and thus enables us to treat arbitrary voltages. (iii) We then study
the transport properties of the renormalized problem using a quench
protocol where we abruptly switch on the impurity-lead coupling
and compute the subsequent time evolution of the current, $J(t)$,
using the time-dependent density-matrix renormalization group (tDMRG)
\cite{Vidal04,Daley04,White04,Schollwoeck11}.  Whereas similar
protocols \cite{Meisner09,Boulat08,Branschadel10,daSilva08} typically
work at $T=0$ 
we consider nonequilibrium thermal leads for
arbitrary 
$T$, using the thermofield approach
\cite{Takahashi75,Barnett87,Das00,deVega15,Guo17}  to describe them with a pure product
state in an enlarged Hilbert space.

We benchmark our approach using the IRLM, finding excellent agreement with exact Bethe-Ansatz predictions for the NESS current. We then turn to the SIAM. For the linear conductance we reproduce equilibrium NRG results. For the nonlinear conductance, we study 
the evolution of the zero-bias peak with $T$ and magnetic field.

\textit{Setup.}--- We consider impurities coupled to two thermal
leads, labeled \mbox{$\alpha\in\{L,R\}$} and characterized by Fermi
functions
$f_\alpha(\omega)=\left(\e^{(\omega-\mu_\alpha)/T}+1\right)^{-1}$,
where \mbox{$\mu_{L/R}=\pm V/{2}$}. (We set
  $e = \hbar = k_{\rm B} = 1$.)
We study two different
 models, the spinless IRLM with a
three-site impurity and Coulomb repulsion~$U$ between neighboring
sites, and the SIAM with Coulomb repulsion $U$ between different spins
and a Zeeman splitting due to a magnetic field $B$. The impurities of
these models are described by
\begin{align}
\notag H_\text{imp}^{\text{(I)}}= &\varepsilon_d\, \hat{n}_C+ U\left(\hat{n}_L+\hat{n}_R-1\right)\hat{n}_C\\
&+\left(t'\, d_C^\dagger d_L^\pdag + t'\, d_C^\dagger d_R^\pdag + \text{H.c.}\right)\\
H_\text{imp}^{\text{(S)}}=&\varepsilon_d \left(\hat{n}_{d\up}+\hat{n}_{d\down}\right) + U\, \hat{n}_{d\up} \hat{n}_{d\down} - \tfrac{B}{2}\left(\hat{n}_{d\up}-\hat{n}_{d\down}\right),
\end{align}
where $\hat{n}_i=d^\dagger_id^\pdag_i,$ for 
$i\in\{L,R,C,d\! \up,d\!\down\}$. In this paper, we focus on the particle-hole symmetric case ($\varepsilon_d=0$ for the IRLM and $\varepsilon_d=-\frac{U}{2}$ for the SIAM).  
The leads are assumed to be noninteracting,
\begin{align}
\label{eq: H_lead}H_\text{lead}^{(\text{I/S})}=&\sum_{\alpha (\sigma) k}\varepsilon^\pdag_{k}c_{\alpha (\sigma) k}^\dagger c_{\alpha (\sigma) k}^\pdag
\equiv
\sum_{q}\varepsilon_q^\pdag c_{q}^\dagger c_{q}^\pdag\,,
\end{align}
with spin index $\sigma\in\left\{\up,\down\right\}$
for the SIAM, 
$q \equiv 
\{\alpha,(\sigma),k\}$ 
a composite index, 
and $k$ a label for 
the energy levels.
The impurity-leads hybridization is given by
\begin{align}
\label{Hhyb}
H_\text{hyb}^{\text{(I/S)}}=
&\sum_{q} \left(v^\pdag_{q}
d^\dagger_{\alpha/\sigma}
c^\pdag_{q}+\text{H.c.}\right)\,,
\end{align}
where in the IRLM the left (right) impurity site $d_L$ ($d_R$) couples to the modes $c_{Lk}$ ($c_{Rk}$), 
respectively, 
while in the SIAM the two spin states $d_\sigma$ couple to the lead modes $c_{\alpha\sigma k}$ spin-independently, $v_{q}=v_{\alpha k}$.
The couplings $v_q$ induce an impurity-lead hybridization $\Gamma_\alpha(\omega)=\pi \sum_{k\sigma} |v_q|^2\delta(\omega-\varepsilon_q)$, 
chosen such that they represent a box distribution $\Gamma_\alpha(\omega)=\Gamma_{\alpha}\Theta(D-|\omega|)$ in the continuum limit with half-bandwidth $D:=1$
set as the unit of energy unless specified otherwise.
For the IRLM we set $\Gamma_L=\Gamma_R=0.5D$ corresponding to the hopping element of a tight-binding chain with half-bandwidth $D$, and for the SIAM we likewise choose $\Gamma_L=\Gamma_R$ and define the total hybridization $\Gamma=\Gamma_L+\Gamma_R$.

\begin{figure}
	\includegraphics[width=\linewidth]{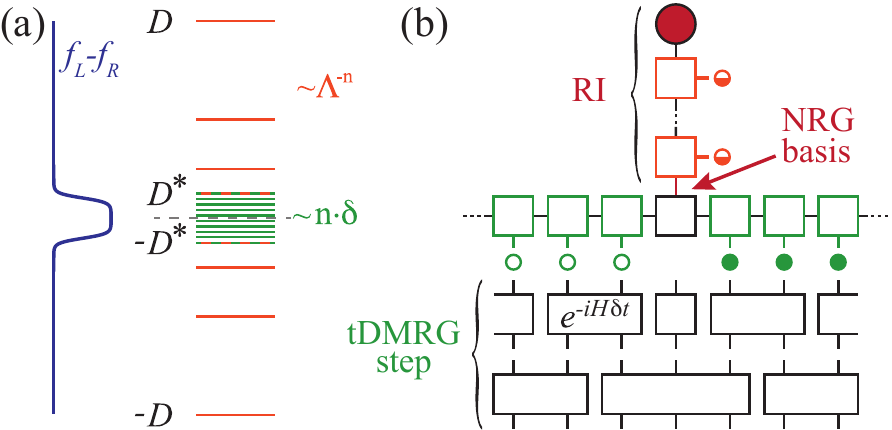}
	\caption{(a) The discretization combines a \mbox{log-sector} for high energy excitations with a lin-sector for the TW.  (b) The \mbox{log-sector} is treated using NRG. Here,  ``holes'' and ``particles'' are recombined. The effective low-energy basis of NRG is used as the local state space of one MPS chain element.
		 For the lin-sector ``holes'' (empty at $t=0$) and ``particles'' (filled at $t=0$) are treated separately. On the chain including the RI, we do a tDMRG calculation based on a Trotter decomposition in ``odd'' and ``even'' bonds \cite{sup}. }
	\label{fig: Fig_DiscretizationAndMPS}
\end{figure}

\textit{Strategy.}---
We describe the thermal leads decoupled from the impurity
using the thermofield approach \cite{Takahashi75,Barnett87,Das00,deVega15}. The impurity-lead coupling induces
nonequilibrium processes, which occur on energy scales corresponding to the \textit{transport window} (TW), defined as the energy range in which $f_L(\omega)\not\approx f_R(\omega)$. Energy scales far outside this TW are effectively in equilibrium and we therefore integrate them  out using NRG, whereas we describe the nonequilibrium physics within the TW using a tDMRG quench. 
We implement both NRG and tDMRG using MPS techniques. 
We use a logarithmically discretized sector (log-sector) representing the energy range of the leads outside the TW and a linearly discretized sector (lin-sector) within the TW, as depicted in Fig.\ \ref{fig: Fig_DiscretizationAndMPS}(a). The transition from the logarithmic to the linear discretization can be smoothened \cite{sup}\nocite{Campo05,Zitko09,Wb11_rho,Wb12_FDM,Wb12_SUN,Corboz10,Schneider06,Wang10,Barthel09,Schmitteckert10,Hanl14}.  To simplify the MPS calculation, we map the leads onto a chain, with on-site and nearest-neighbor terms only, by tridiagonalizing the Hamiltonian. 
Integrating out the log-sector using NRG we get a
\textit{renormalized impurity} (RI) \cite{Guttge2013} and a reduced
effective bandwidth, $2D^*$, of order of the size of the TW.  This
enables us to treat transport on energy scales much smaller than
$D$. In particular, we can study arbitrary ratios of $V/T_K$ in the
SIAM, even if $T_K\ll D$.  We then turn on the coupling between the
log-sector and lin-sector by performing a tDMRG quench, starting from
an initial state
$\ket{\Psi_\text{ini}}=\ket{\phi_\text{ini}}\otimes\ket{\Omega_\text{lin}}$,
where $\ket{\phi_\text{ini}}$ describes the initial state of the RI,
and $\ket{\Omega_\text{lin}}$ is a \textit{pure product state}
describing the lin-sector of the thermal leads in the thermofield
approach.  To describe steady-state properties, we time-evolve
$\ket{\Psi_\text{ini}}$ until expectation values are stationary up to
oscillations around their mean value.  Since the effective bandwidth
relevant for this tDMRG calculation is given by $D^*$, not $D$,
exponentially large time scales of order $1/D^*\gg1/D$ are accessible.

\textit{Thermofield description of decoupled leads.}---
In the context of MPS methods the thermofield description \cite{Takahashi75,Barnett87,Das00,deVega15} of the decoupled leads has two advantages:
finite temperature states are represented as pure states, and thermal leads are described by a simple product state. 

Akin to purification \cite{Schollwoeck11}
we double our Hilbert space by introducing one auxiliary mode ${c}_{q2}$ (not coupled to the system) for each lead mode $c_{q1}=c_{q}$.  In this enlarged Hilbert space we define a pure state $\ket{\Omega}$ such that the thermal expectation value of an operator $A$ acting on the original physical lead is given by $\braket{A}=\braket{\Omega|A|\Omega}$.
This state can be written as \cite{sup} 
\begin{align}
\label{eq: Omega}
\ket{\Omega}&=\prod_q
\underbrace{\bigl(\sqrt{1-f_q}\ket{0,1}_q+\sqrt{f_q}\ket{1,0}_q\bigr)}_{ \equiv \, \ket{\tilde{0},\tilde{1}}_q}
\,,
\end{align}
with $f_q=f_\alpha(\varepsilon_q)$, where $\ket{0,1}_q$ and $\ket{1,0}_q$  are defined by \mbox{$c_{q1}^\pdag\ket{0,1}_q$} \mbox{$=c_{q2}^\dagger\ket{0,1}_q$} \mbox{$=c_{q1}^\dagger\ket{1,0}_q$} \mbox{$=c_{q2}^\pdag\ket{1,0}_q=0$} for all $q$.
We map $\ket{\Omega}$ to a pure product state using the rotation 
\begin{align}
\label{eq: rotation_TF}\begin{pmatrix}\tilde{c}_{q1}\\\tilde{c}_{q2}\end{pmatrix}=\begin{pmatrix*}
\sqrt{1-f_q} & -\sqrt{f_q} \\\sqrt{f_q} & \sqrt{1-f_q} \end{pmatrix*}\begin{pmatrix}{c}_{q1}\\{c}_{q2}\end{pmatrix}
\text{.}
\end{align}
Having $\tilde{c}_{q1}^\pdag\ket{\Omega}
=\tilde{c}_{q2}^\dagger\ket{\Omega}=0$, the modes
$\tilde{c}_{q1}$ ($\tilde{c}_{q2}$) can 
be interpreted as ``holes'' (``particles'') which are empty (filled) in the thermal state, respectively.
Since in Eq.~(\,\ref{eq: Omega}) we constructed $\ket{\Omega}$
to be an eigenstate of the particle number operator,
it remains so in the rotated basis.
The physical and auxiliary
modes are decoupled in the
unrotated basis, hence we are free to choose an arbitrary Hamiltonian
(and hence time evolution) for
the auxiliary modes \cite{Karrasch12}.
We choose 
their single-particle energies 
equal to those of the physical modes, 
$\varepsilon_{q2} = \varepsilon_{q}$,  
in order to ensure that the resulting total lead Hamiltonian
is diagonal in $j$ in both the original \textit{and}
the rotated basis:
\begin{align}
\mathcal{H}_\mathrm{lead} \equiv
H_\text{lead}+{H}_\text{aux}
=& \sum_{qj} \varepsilon_q{c}_{qj}^\dagger{c}_{qj}^\pdag
\!=  \sum_{qj}\varepsilon_q\tilde{c}_{qj}^\dagger
\tilde{c}_{qj}^\pdag
\, . \label{eq:Haux}
\end{align}%
\Eq{Hhyb} %
is rotated into 
$H_\text{hyb}^{(\text{I/S})} =
\sum_{qj} \bigl( \tilde{v}_{qj}d_{\alpha/\sigma}^\dagger
\tilde c_{qj}+\text{H.c.}\bigr)$
whose 
couplings $\tilde{v}_{q1}=v_{q}\sqrt{1-f_q}$ and $\tilde{v}_{q2}=v_{q}\sqrt{f_q}$,
now {\it explicitly depend}
on the Fermi function
and encode all relevant information about
temperature and voltage. 

For the SIAM, we use a specific linear combination of $\tilde{c}_{Lk\sigma i}$ and $\tilde{c}_{Rk\sigma i}$ modes,
$\tilde{C}_{k\sigma i}\propto\sum_\alpha \tilde{v}_{\alpha k \sigma i}\tilde{c}_{\alpha k \sigma i}$, because the modes orthogonal to these \cite{sup} decouple.
Mixing left and right lead modes is possible
despite the nonequilibrium situation 
because the difference in chemical potentials is accounted for by the $V$-dependent couplings~$\tilde{v}_q$. In the IRLM this reduction of modes is not possible because left and right lead couple to different impurity sites.

\textit{NRG renormalization of the impurity.}--- As is standard
  for NRG we map the leads (in the thermofield representation) from
  the original ``star geometry'' to a chain geometry. To ensure that
  $\ket{\Omega}$ remains a product state, we perform the
corresponding unitary transformation for ``holes'' and ``particles''
independently.  This results in a chain consisting of two channels
$i\in\{1,2\}$ for the SIAM, and four for the IRLM due to the
additional lead index $\alpha\in\{L,R\}$. The first part of the chain
corresponds to the log-sector, the later part to the lin-sector.  The
hoppings within the log-sector decay as $\Lambda^{-n}$, because for
each lead level~$q$ within the \mbox{log-sector} of the original star
geometry either $\tilde{c}_{q1}$ or $\tilde{c}_{q2}$ decouples from
the RI, due to $f_q\in\{0,1\}$.  For NRG calculations it is
unfavorable to describe ``holes'' and ``particles'' in separate
chains, because then particle-hole excitations involve opposite levels
of different chains. For that reason we recombine the ``holes'' and
``particles'' of the \mbox{log-sector} into one chain using a further
tridiagonalization. In the IRLM this is done for each lead $\alpha$
independently. After that, the log-sector resembles a standard Wilson
chain with hoppings that scale as $\Lambda^{-n/2}$, reflecting the
fact that the \mbox{log-sector} is effectively in equilibrium.  A
sketch of the different geometries can be found in Fig.\ S2 of
Ref.~\cite{sup}.

Using NRG, we find an effective low-energy many-body basis for the log-sector, which we interpret as the local state space of a RI and treat it as one chain element of our MPS chain. Coupled to this RI we have the lin-sector of the leads, represented as two separate chains for ``holes'' and ``particles'', as shown in the upper part of Fig.\ \ref{fig: Fig_DiscretizationAndMPS}(b).

\textit{tDMRG quench.}---
We choose the initial state for the quench as the product state $\ket{\Psi_\text{ini}}=\ket{\phi_\text{ini}}\otimes\ket{\Omega_\text{lin}}$\,. This implies that for the lin-sector we start with the state in which all ``holes'' (``particles'') are empty (filled).
As the initial state of the RI, $\ket{\phi_\text{ini}}$,
we choose a ground state of the NRG basis 
(in principle one can choose any of the low-energy basis
states whose excitation energy is well within the TW). 
We then switch on the coupling between the RI and the leads smoothly
over a short time window.  The system time-evolves under the
Hamiltonian
$\hat{H}=H_\text{imp}+H_\text{hyb}+H_\text{lead}+{H}_\text{aux}$,
$\ket{\Psi(t)}=\e^{-\ii\hat{H}t}\ket{\Psi_\text{ini}}$.  We perform
the time-evolution using tDMRG based on a second order Trotter
decomposition, as depicted in Fig.\,\ref{fig:
  Fig_DiscretizationAndMPS}(b), with a Trotter time step of order
$1/D^*$. (Technical details can be found in section~S-3.C of
  Ref.~\cite{sup}.)  The fact that this initial lead state is
entanglement-free is advantageous for reaching comparatively long
times. We extract NESS information from
$\braket{A(t)} = \braket{\Psi(t)|A|\Psi(t)}$ within a window of
intermediate times, large enough for post-quench transients to no
longer dominate, but well below the recurrence time, where finite-size
effects set in.  We compute the current through the impurity site
(SIAM) or the central impurity site (IRLM), respectively, using
$J=\frac{1}{2}(J_L-J_R)$, where $J_L$ ($J_R$) is the current that
flows into the site from the left (right), respectively \cite{sup}.
We are able to track the time evolution up to times of order
  $1/D^*$. Since $D^*\sim\max(V,T)$, this suffices to describe particle
  transport for any choice of $V$ or $T$. However, processes on much
  smaller energy scales cannot necessarily be resolved (see 
  section~S-4.C of \cite{sup} for details).

\begin{figure}[t]
\includegraphics[width=\linewidth]{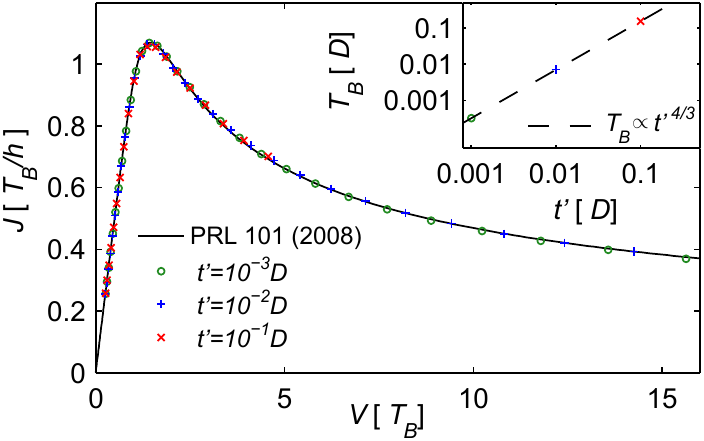}
\vspace{-0.5cm}
\caption{
Universal scaling of current vs. voltage
for the IRLM at the self-dual point for $T=0$
in units of the energy scale $T_B(t')$
with negative
differential conductance at large voltages,
in excellent agreement with analytical results 
(solid curve,
\cite{Boulat08}). The inset shows the scaling of
$T_B$ with $(t')^{3/4}$.
}
\label{fig: IRLM Scaling}
\vspace{-0.35cm}
\end{figure}

\textit{Interacting Resonant Level Model.}---
We benchmark our method
for the IRLM, for which Ref.\ \cite{Boulat08} 
computed the steady-state current at $T=0$
both numerically using DMRG quenches and analytically
using the exact Bethe ansatz. A universal scaling of
the current-voltage characteristics was found at the
self-dual point of the model, with the corresponding
energy scale $T_B $ scaling as
$(t')^{3/4}$. (These 
results were very recently confirmed by Ref.\ \cite{Bidzhiev17}.) 
Fig.\,\ref{fig: IRLM Scaling} presents a comparison of 
our data with the analytical expression for the
universal scaling curve
 given in  \cite{Boulat08},
for
the current as function of voltage at $T=0$ at
the self-dual point $U\approx D$ and $\varepsilon_d=0$.
The agreement is excellent  for a large range
of $t'$ values. For each value of $t'$, $T_B$ was used
as a fit parameter; the resulting $T_B$~values,
shown in the inset, agree nicely with the scaling
predicted in  [\onlinecite{Boulat08}].
Using the fitted values of $T_B$, all data points deviate by less than 2\% from the Bethe results. 
Our use of NRG to renormalize the impurity
enables us
to study values of $t'$ up to a hundred times smaller
than the values
 used
in \cite{Boulat08}, giving us
 access much
 smaller 
values of $T_B$
and
larger $V/T_B$~ratios.

\textit{Single-Impurity Anderson Model.}---
\begin{figure*}[]
	\centering
\includegraphics[width=\linewidth]{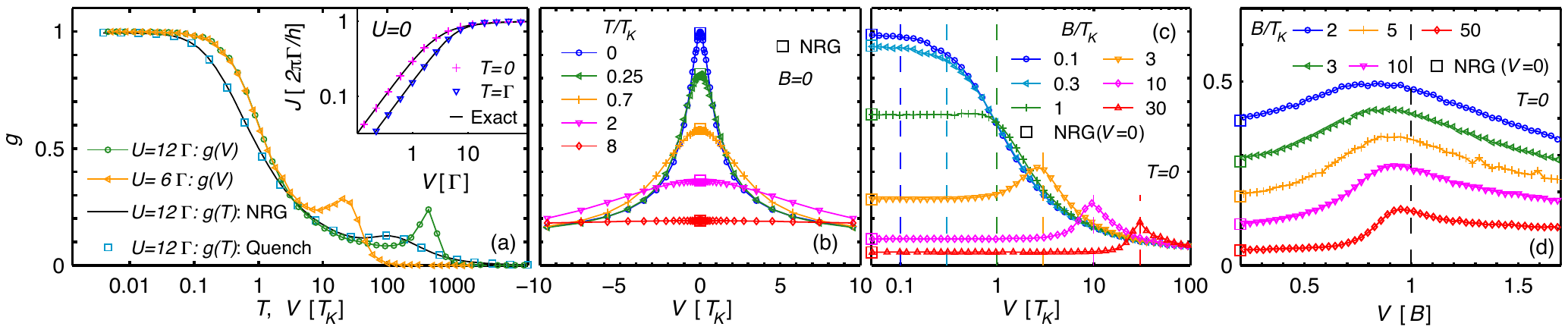}

\vspace{-0.7\baselineskip} 
\caption{
Numerical results for the SIAM with $\Gamma=10^{-3}$.
For $U=12\Gamma$, used in (b-d), we find
$T_K = 2.61\cdot 10^{-5}$. (This implies $T_K=1.04T_K^{(\chi)}$, where $T_K^{(\chi)}=\frac{1}{4\chi_s}=(U\Gamma/2)^{\frac{1}{2}}e^{\pi\left(\frac{\Gamma}{2U}-\frac{U}{8\Gamma}\right)}$ is an alternative definition of the Kondo temperature based on the Bethe-Ansatz result \cite{Wiegmann83,*Tsvelick83} for the static spin susceptibility $\chi_s$, at $B=T=0.$)
(a) Conductance vs.\ 
$V$ and $T$: squares show quench results
in linear response as function of $T$,
$g(T,0)$, in good agreement with NRG results
(solid line). Dots and triangles show quench results
for the nonlinear conductance vs.\ 
$V$ at $T=0$ for two different values of $U$.
Inset: current vs.\ 
$V$ for $U=0$ on a log-log scale, for two different temperatures,
showing excellent
agreement with analytical results. 
(b) Disappearance of the Kondo resonance in $g(T,V)$
with increasing $T$ at $B=0$,
with $g(T,-V)=g(T,V)$, by symmetry.
(c) Splitting of the resonance in $g(0,V)$
for finite
 $B$.
Two subpeaks emerge
at $V\approx\pm B$, as marked by the dashed lines.
(d) Similar data as in (c) but plotted vs.\ $V/B$ and
on a linear scale.
For $B=2T_K$ the peak position in the conductance $g(0,V)$
is still slightly below $B$, but for higher magnetic field the peak clearly moves towards $V/B\approx1$. 
In (b)-(d) the squares indicate the NRG result for $V=0$.	
	}
	\label{fig: GoverTV} \vspace{-1.1\baselineskip}
\end{figure*}
For the SIAM, a natural first check is the noninteracting case, $U=0$,
which is exactly solvable, but its treatment in MPS numerics does not
differ from the case $U\neq0$.  The inset of Fig.\,\ref{fig:
  GoverTV}(a) displays the current over voltage for two different
temperatures, showing good agreement between our MPS numerics and
exact predictions, thus providing direct evidence for the validity of
our approach.  For $U\neq0$, our method yields quantitative
  agreement with previous numerical results obtained in the regime
  $V\gtrsim\Gamma$ \cite{Eckel10,Werner10}, see 
  section~S-6 of Ref.~\cite{sup} 
  for details. Furthermore, we find good 
  agreement with the auxiliary master equation approach for arbitrary
  voltages, see Ref.\ \cite{Fugger17} for details.

The main panel of Fig.\,\ref{fig: GoverTV}(a) focuses
on the differential conductance $g(T,V) =
\tfrac{\partial J(T,V)}{\partial V} /
\tfrac{2e^2}{h}$ 
for strong interactions.  As a consistency check,
we compare our results for $g(T,0)$ with
the linear conductance computed using
FDM-NRG \cite{AW07}.
We find excellent agreement over a large
range of temperatures. From this data, we define
the Kondo temperature $T_K$ via the condition
$\ds{g(T_K,0)\equiv\tfrac{1}{2}}$.

We also show 
$\ds{g(0,V)}$ over a wide 
voltage range in Fig.\,\ref{fig: GoverTV}(a).
In agreement with experiment \cite{Kretinin12} 
and other theoretical work \cite{Pletyukhov12}
this curve lies above $\ds{g(T,0)}$. 
The difference can be quantified by the value of $g(0,T_K)$,
a universal number characterizing NESS transport for the
SIAM, whose precise value is not yet known with
quantitative certainty. Our method, which we trust
to be quantitatively reliable, yields $g(0,T_K)
\approx 0.60 \pm 0.02$
in the Kondo limit of $U/\Gamma\gg1$,
where the estimated error bar of about 3\%
is likely conservative (cf.\ \cite{sup}).
For comparison, (nonexact) analytical calculations
for the Kondo model yielded $g(0,T_K)\approx 2/3$ \cite{Pletyukhov12,Smirnov13}. 

Fig.~\ref{fig: GoverTV}(b-d) show our quantitative description of the $T$- and $B$-dependence of the zero-bias peak in the Kondo limit ($U/\Gamma=12$). With increasing $T$ at $B=0$, the zero-bias peak decreases [Fig.~\ref{fig: GoverTV}(b)], as observed in numerous experiments \cite{Goldhaber98,Cronenwett98,Simmel99,vanderWiel00,Kretinin11,Kretinin12}.
For finite $B$,
the zero-bias peak
splits into two sub-peaks at  $\ds{V\approx\pm B}$ [Fig. \ref{fig: GoverTV}(c)].
A more detailed analysis of the value of $B$ at which the peak begins to split \cite{Filippone18,
Oguri17,*Oguri17_2,*Oguri17_3} is given in  
 section~S-7 of Ref.~\cite{sup}.
 In Fig.\ \ref{fig: GoverTV}(d)
the peak position with respect to $B$
is
resolved in more detail, with the voltage given in
units of $B$. While for $B\approx 2T_K$ the peak position
is roughly at  $V/B \approx 0.83$, it
quickly tends towards 
$V/B=1$ for larger magnetic fields. Our study thus quantitatively confirms that the large-field peak-to-peak splitting for the nonlinear conductance is $\approx2B$, as observed in several experiments \cite{Ralph94,Goldhaber98, vanderWiel00}. 
This is also found in 
 independent calculations \cite{Fugger17} using the approach of Ref.~[\onlinecite{Dorda15}].

\textit{Summary and Outlook.}--- We have combined the thermofield
approach with a hybrid NRG-tDMRG quench strategy to reach a
longstanding goal: a versatile, flexible, and \textit{quantitatively
  reliable} method for studying quantum impurity models in
steady-state nonequilibrium.  Due to these features,
our scheme has the potential of developing into the
  method of choice for such settings, in the same way as NRG is the
  method of choice for equilibrium impurity models. Indeed, various
  quantitative benchmark tests have confirmed the accuracy of our
  scheme, and it can easily be applied to other models and
  setups. For example, a generalization to a finite temperature
difference between left and right lead would be straightforward.  It
would also be interesting to use our setup for quantitative studies of
the nonequilibrium two-channel Kondo physics measured in
\cite{Iftikhar15}, or to study impurity models with superconducting
leads, since the hybrid NRG-tDMRG approach is ideally suited for
dealing with the bulk gap.

Methodologically, our setup can straightforwardly be extended to study NESS physics without resorting to a quench strategy by including Lindblad driving terms in the Liouville equation, which are \textit{local} on the MPS chain \cite{Schwarz16}. Although the direct time-evolution of such Lindblad equations based on tensor networks seems feasible \cite{Werner16}, one could try to avoid the real-time evolution
altogether and target the steady-state directly by looking for the density matrix which fulfills $\dot{\rho}=0$ \cite{Cui15,Mascarenhas15}.

We thank F.\,Heidrich-Meisner and P.\,Werner for providing the
  reference data in Fig.~S6. We acknowledge useful discussions with
E.\,Arrigoni, M.-C.\,Ba\~nuls, B.\,Bruognolo, A.\,Dorda, D.\,Fugger,
M.\,Goldstein and H.\,Sch\"oller. This work was supported by the
German-Israeli-Foundation through I-1259-303.10 and by the DFG through
the excellence cluster NIM. A.\,W.\ was also supported by WE4819/1-1
and WE4819/2-1. I.\,W.\ was supported by National Science Centre in
Poland through the Project No.\ DEC-2013/10/E/ST3/00213.

\bibliography{Quenches_bib} 

\begin{thebibliography}{64}%
\makeatletter
\providecommand \@ifxundefined [1]{%
 \@ifx{#1\undefined}
}%
\providecommand \@ifnum [1]{%
 \ifnum #1\expandafter \@firstoftwo
 \else \expandafter \@secondoftwo
 \fi
}%
\providecommand \@ifx [1]{%
 \ifx #1\expandafter \@firstoftwo
 \else \expandafter \@secondoftwo
 \fi
}%
\providecommand \natexlab [1]{#1}%
\providecommand \enquote  [1]{``#1''}%
\providecommand \bibnamefont  [1]{#1}%
\providecommand \bibfnamefont [1]{#1}%
\providecommand \citenamefont [1]{#1}%
\providecommand \href@noop [0]{\@secondoftwo}%
\providecommand \href [0]{\begingroup \@sanitize@url \@href}%
\providecommand \@href[1]{\@@startlink{#1}\@@href}%
\providecommand \@@href[1]{\endgroup#1\@@endlink}%
\providecommand \@sanitize@url [0]{\catcode `\\12\catcode `\$12\catcode
  `\&12\catcode `\#12\catcode `\^12\catcode `\_12\catcode `\%12\relax}%
\providecommand \@@startlink[1]{}%
\providecommand \@@endlink[0]{}%
\providecommand \url  [0]{\begingroup\@sanitize@url \@url }%
\providecommand \@url [1]{\endgroup\@href {#1}{\urlprefix }}%
\providecommand \urlprefix  [0]{URL }%
\providecommand \Eprint [0]{\href }%
\providecommand \doibase [0]{http://dx.doi.org/}%
\providecommand \selectlanguage [0]{\@gobble}%
\providecommand \bibinfo  [0]{\@secondoftwo}%
\providecommand \bibfield  [0]{\@secondoftwo}%
\providecommand \translation [1]{[#1]}%
\providecommand \BibitemOpen [0]{}%
\providecommand \bibitemStop [0]{}%
\providecommand \bibitemNoStop [0]{.\EOS\space}%
\providecommand \EOS [0]{\spacefactor3000\relax}%
\providecommand \BibitemShut  [1]{\csname bibitem#1\endcsname}%
\let\auto@bib@innerbib\@empty
\bibitem [{\citenamefont {Rosch}\ \emph {et~al.}(2003)\citenamefont {Rosch},
  \citenamefont {Paaske}, \citenamefont {Kroha},\ and\ \citenamefont
  {W\"olfle}}]{Rosch03}%
  \BibitemOpen
  \bibfield  {author} {\bibinfo {author} {\bibfnamefont {A.}~\bibnamefont
  {Rosch}}, \bibinfo {author} {\bibfnamefont {J.}~\bibnamefont {Paaske}},
  \bibinfo {author} {\bibfnamefont {J.}~\bibnamefont {Kroha}}, \ and\ \bibinfo
  {author} {\bibfnamefont {P.}~\bibnamefont {W\"olfle}},\ }\bibfield  {title}
  {\enquote {\bibinfo {title} {Nonequilibrium transport through a {Kondo} dot
  in a magnetic field: Perturbation theory and poor man's scaling},}\ }\href
  {\doibase 10.1103/PhysRevLett.90.076804} {\bibfield  {journal} {\bibinfo
  {journal} {Phys. Rev. Lett.}\ }\textbf {\bibinfo {volume} {90}},\ \bibinfo
  {pages} {076804} (\bibinfo {year} {2003})}\BibitemShut {NoStop}%
\bibitem [{\citenamefont {Kehrein}(2005)}]{Kehrein05}%
  \BibitemOpen
  \bibfield  {author} {\bibinfo {author} {\bibfnamefont {S.}~\bibnamefont
  {Kehrein}},\ }\bibfield  {title} {\enquote {\bibinfo {title} {Scaling and
  decoherence in the nonequilibrium {Kondo} model},}\ }\href {\doibase
  10.1103/PhysRevLett.95.056602} {\bibfield  {journal} {\bibinfo  {journal}
  {Phys. Rev. Lett.}\ }\textbf {\bibinfo {volume} {95}},\ \bibinfo {pages}
  {056602} (\bibinfo {year} {2005})}\BibitemShut {NoStop}%
\bibitem [{\citenamefont {Anders}(2008)}]{Anders08}%
  \BibitemOpen
  \bibfield  {author} {\bibinfo {author} {\bibfnamefont {F.~B.}\ \bibnamefont
  {Anders}},\ }\bibfield  {title} {\enquote {\bibinfo {title} {Steady-state
  currents through nanodevices: A scattering-states numerical
  renormalization-group approach to open quantum systems},}\ }\href {\doibase
  10.1103/PhysRevLett.101.066804} {\bibfield  {journal} {\bibinfo  {journal}
  {Phys. Rev. Lett.}\ }\textbf {\bibinfo {volume} {101}},\ \bibinfo {pages}
  {066804} (\bibinfo {year} {2008})}\BibitemShut {NoStop}%
\bibitem [{\citenamefont {Kirino}\ \emph {et~al.}(2008)\citenamefont {Kirino},
  \citenamefont {Fujii}, \citenamefont {Zhao},\ and\ \citenamefont
  {Ueda}}]{Kirino08}%
  \BibitemOpen
  \bibfield  {author} {\bibinfo {author} {\bibfnamefont {S.}~\bibnamefont
  {Kirino}}, \bibinfo {author} {\bibfnamefont {T.}~\bibnamefont {Fujii}},
  \bibinfo {author} {\bibfnamefont {J.}~\bibnamefont {Zhao}}, \ and\ \bibinfo
  {author} {\bibfnamefont {K.}~\bibnamefont {Ueda}},\ }\bibfield  {title}
  {\enquote {\bibinfo {title} {Time-dependent {DMRG} study on quantum dot under
  a finite bias voltage},}\ }\href {\doibase 10.1143/JPSJ.77.084704} {\bibfield
   {journal} {\bibinfo  {journal} {Journal of the Physical Society of Japan}\
  }\textbf {\bibinfo {volume} {77}},\ \bibinfo {pages} {084704} (\bibinfo
  {year} {2008})}\BibitemShut {NoStop}%
\bibitem [{\citenamefont {Heidrich-Meisner}\ \emph {et~al.}(2009)\citenamefont
  {Heidrich-Meisner}, \citenamefont {Feiguin},\ and\ \citenamefont
  {Dagotto}}]{Meisner09}%
  \BibitemOpen
  \bibfield  {author} {\bibinfo {author} {\bibfnamefont {F.}~\bibnamefont
  {Heidrich-Meisner}}, \bibinfo {author} {\bibfnamefont {A.~E.}\ \bibnamefont
  {Feiguin}}, \ and\ \bibinfo {author} {\bibfnamefont {E.}~\bibnamefont
  {Dagotto}},\ }\bibfield  {title} {\enquote {\bibinfo {title} {Real-time
  simulations of nonequilibrium transport in the single-impurity {Anderson}
  model},}\ }\href {\doibase 10.1103/PhysRevB.79.235336} {\bibfield  {journal}
  {\bibinfo  {journal} {Phys. Rev. B}\ }\textbf {\bibinfo {volume} {79}},\
  \bibinfo {pages} {235336} (\bibinfo {year} {2009})}\BibitemShut {NoStop}%
\bibitem [{\citenamefont {Eckel}\ \emph {et~al.}(2010)\citenamefont {Eckel},
  \citenamefont {Heidrich-Meisner}, \citenamefont {Jakobs}, \citenamefont
  {Thorwart}, \citenamefont {Pletyukhov},\ and\ \citenamefont
  {Egger}}]{Eckel10}%
  \BibitemOpen
  \bibfield  {author} {\bibinfo {author} {\bibfnamefont {J.}~\bibnamefont
  {Eckel}}, \bibinfo {author} {\bibfnamefont {F.}~\bibnamefont
  {Heidrich-Meisner}}, \bibinfo {author} {\bibfnamefont {S.~G.}\ \bibnamefont
  {Jakobs}}, \bibinfo {author} {\bibfnamefont {M.}~\bibnamefont {Thorwart}},
  \bibinfo {author} {\bibfnamefont {M.}~\bibnamefont {Pletyukhov}}, \ and\
  \bibinfo {author} {\bibfnamefont {R.}~\bibnamefont {Egger}},\ }\bibfield
  {title} {\enquote {\bibinfo {title} {Comparative study of theoretical methods
  for non-equilibrium quantum transport},}\ }\href
  {http://stacks.iop.org/1367-2630/12/i=4/a=043042} {\bibfield  {journal}
  {\bibinfo  {journal} {New J. Phys.}\ }\textbf {\bibinfo {volume} {12}},\
  \bibinfo {pages} {043042} (\bibinfo {year} {2010})}\BibitemShut {NoStop}%
\bibitem [{\citenamefont {Werner}\ \emph {et~al.}(2010)\citenamefont {Werner},
  \citenamefont {Oka}, \citenamefont {Eckstein},\ and\ \citenamefont
  {Millis}}]{Werner10}%
  \BibitemOpen
  \bibfield  {author} {\bibinfo {author} {\bibfnamefont {P.}~\bibnamefont
  {Werner}}, \bibinfo {author} {\bibfnamefont {T.}~\bibnamefont {Oka}},
  \bibinfo {author} {\bibfnamefont {M.}~\bibnamefont {Eckstein}}, \ and\
  \bibinfo {author} {\bibfnamefont {A.~J.}\ \bibnamefont {Millis}},\ }\bibfield
   {title} {\enquote {\bibinfo {title} {Weak-coupling quantum {Monte Carlo}
  calculations on the {Keldysh} contour: {Theory} and application to the
  current-voltage characteristics of the {Anderson} model},}\ }\href {\doibase
  10.1103/PhysRevB.81.035108} {\bibfield  {journal} {\bibinfo  {journal} {Phys.
  Rev. B}\ }\textbf {\bibinfo {volume} {81}},\ \bibinfo {pages} {035108}
  (\bibinfo {year} {2010})}\BibitemShut {NoStop}%
\bibitem [{\citenamefont {Pletyukhov}\ and\ \citenamefont
  {Schoeller}(2012)}]{Pletyukhov12}%
  \BibitemOpen
  \bibfield  {author} {\bibinfo {author} {\bibfnamefont {M.}~\bibnamefont
  {Pletyukhov}}\ and\ \bibinfo {author} {\bibfnamefont {H.}~\bibnamefont
  {Schoeller}},\ }\bibfield  {title} {\enquote {\bibinfo {title}
  {Nonequilibrium {Kondo} model: Crossover from weak to strong coupling},}\
  }\href {\doibase 10.1103/PhysRevLett.108.260601} {\bibfield  {journal}
  {\bibinfo  {journal} {Phys. Rev. Lett.}\ }\textbf {\bibinfo {volume} {108}},\
  \bibinfo {pages} {260601} (\bibinfo {year} {2012})}\BibitemShut {NoStop}%
\bibitem [{\citenamefont {Smirnov}\ and\ \citenamefont
  {Grifoni}(2013)}]{Smirnov13}%
  \BibitemOpen
  \bibfield  {author} {\bibinfo {author} {\bibfnamefont {S.}~\bibnamefont
  {Smirnov}}\ and\ \bibinfo {author} {\bibfnamefont {M.}~\bibnamefont
  {Grifoni}},\ }\bibfield  {title} {\enquote {\bibinfo {title} {{Keldysh}
  effective action theory for universal physics in {spin-$\frac{1}{2}$} {Kondo}
  dots},}\ }\href {\doibase 10.1103/PhysRevB.87.121302} {\bibfield  {journal}
  {\bibinfo  {journal} {Phys. Rev. B}\ }\textbf {\bibinfo {volume} {87}},\
  \bibinfo {pages} {121302} (\bibinfo {year} {2013})}\BibitemShut {NoStop}%
\bibitem [{\citenamefont {Cohen}\ \emph {et~al.}(2014)\citenamefont {Cohen},
  \citenamefont {Gull}, \citenamefont {Reichman},\ and\ \citenamefont
  {Millis}}]{Cohen14}%
  \BibitemOpen
  \bibfield  {author} {\bibinfo {author} {\bibfnamefont {G.}~\bibnamefont
  {Cohen}}, \bibinfo {author} {\bibfnamefont {E.}~\bibnamefont {Gull}},
  \bibinfo {author} {\bibfnamefont {D.~R.}\ \bibnamefont {Reichman}}, \ and\
  \bibinfo {author} {\bibfnamefont {A.~J.}\ \bibnamefont {Millis}},\ }\bibfield
   {title} {\enquote {\bibinfo {title} {{Green}'s functions from real-time
  bold-line {Monte Carlo} calculations: Spectral properties of the
  nonequilibrium {Anderson} impurity model},}\ }\href {\doibase
  10.1103/PhysRevLett.112.146802} {\bibfield  {journal} {\bibinfo  {journal}
  {Phys. Rev. Lett.}\ }\textbf {\bibinfo {volume} {112}},\ \bibinfo {pages}
  {146802} (\bibinfo {year} {2014})}\BibitemShut {NoStop}%
\bibitem [{\citenamefont {Antipov}\ \emph {et~al.}(2016)\citenamefont
  {Antipov}, \citenamefont {Dong},\ and\ \citenamefont {Gull}}]{Antipov16}%
  \BibitemOpen
  \bibfield  {author} {\bibinfo {author} {\bibfnamefont {A.~E.}\ \bibnamefont
  {Antipov}}, \bibinfo {author} {\bibfnamefont {Q.}~\bibnamefont {Dong}}, \
  and\ \bibinfo {author} {\bibfnamefont {E.}~\bibnamefont {Gull}},\ }\bibfield
  {title} {\enquote {\bibinfo {title} {Voltage quench dynamics of a {Kondo}
  system},}\ }\href {\doibase 10.1103/PhysRevLett.116.036801} {\bibfield
  {journal} {\bibinfo  {journal} {Phys. Rev. Lett.}\ }\textbf {\bibinfo
  {volume} {116}},\ \bibinfo {pages} {036801} (\bibinfo {year}
  {2016})}\BibitemShut {NoStop}%
\bibitem [{\citenamefont {Reininghaus}\ \emph {et~al.}(2014)\citenamefont
  {Reininghaus}, \citenamefont {Pletyukhov},\ and\ \citenamefont
  {Schoeller}}]{Reininghaus14}%
  \BibitemOpen
  \bibfield  {author} {\bibinfo {author} {\bibfnamefont {F.}~\bibnamefont
  {Reininghaus}}, \bibinfo {author} {\bibfnamefont {M.}~\bibnamefont
  {Pletyukhov}}, \ and\ \bibinfo {author} {\bibfnamefont {H.}~\bibnamefont
  {Schoeller}},\ }\bibfield  {title} {\enquote {\bibinfo {title} {{Kondo} model
  in nonequilibrium: Interplay between voltage, temperature, and crossover from
  weak to strong coupling},}\ }\href {\doibase 10.1103/PhysRevB.90.085121}
  {\bibfield  {journal} {\bibinfo  {journal} {Phys. Rev. B}\ }\textbf {\bibinfo
  {volume} {90}},\ \bibinfo {pages} {085121} (\bibinfo {year}
  {2014})}\BibitemShut {NoStop}%
\bibitem [{\citenamefont {Dorda}\ \emph {et~al.}(2015)\citenamefont {Dorda},
  \citenamefont {Ganahl}, \citenamefont {Evertz}, \citenamefont {von~der
  Linden},\ and\ \citenamefont {Arrigoni}}]{Dorda15}%
  \BibitemOpen
  \bibfield  {author} {\bibinfo {author} {\bibfnamefont {A.}~\bibnamefont
  {Dorda}}, \bibinfo {author} {\bibfnamefont {M.}~\bibnamefont {Ganahl}},
  \bibinfo {author} {\bibfnamefont {H.~G.}\ \bibnamefont {Evertz}}, \bibinfo
  {author} {\bibfnamefont {W.}~\bibnamefont {von~der Linden}}, \ and\ \bibinfo
  {author} {\bibfnamefont {E.}~\bibnamefont {Arrigoni}},\ }\bibfield  {title}
  {\enquote {\bibinfo {title} {Auxiliary master equation approach within matrix
  product states: Spectral properties of the nonequilibrium {Anderson} impurity
  model},}\ }\href {\doibase 10.1103/PhysRevB.92.125145} {\bibfield  {journal}
  {\bibinfo  {journal} {Phys. Rev. B}\ }\textbf {\bibinfo {volume} {92}},\
  \bibinfo {pages} {125145} (\bibinfo {year} {2015})}\BibitemShut {NoStop}%
\bibitem [{\citenamefont {Jakobs}\ \emph {et~al.}(2007)\citenamefont {Jakobs},
  \citenamefont {Meden},\ and\ \citenamefont {Schoeller}}]{Jakobs07}%
  \BibitemOpen
  \bibfield  {author} {\bibinfo {author} {\bibfnamefont {S.~G.}\ \bibnamefont
  {Jakobs}}, \bibinfo {author} {\bibfnamefont {V.}~\bibnamefont {Meden}}, \
  and\ \bibinfo {author} {\bibfnamefont {H.}~\bibnamefont {Schoeller}},\
  }\bibfield  {title} {\enquote {\bibinfo {title} {Nonequilibrium functional
  renormalization group for interacting quantum systems},}\ }\href {\doibase
  10.1103/PhysRevLett.99.150603} {\bibfield  {journal} {\bibinfo  {journal}
  {Phys. Rev. Lett.}\ }\textbf {\bibinfo {volume} {99}},\ \bibinfo {pages}
  {150603} (\bibinfo {year} {2007})}\BibitemShut {NoStop}%
\bibitem [{\citenamefont {Boulat}\ \emph {et~al.}(2008)\citenamefont {Boulat},
  \citenamefont {Saleur},\ and\ \citenamefont {Schmitteckert}}]{Boulat08}%
  \BibitemOpen
  \bibfield  {author} {\bibinfo {author} {\bibfnamefont {E.}~\bibnamefont
  {Boulat}}, \bibinfo {author} {\bibfnamefont {H.}~\bibnamefont {Saleur}}, \
  and\ \bibinfo {author} {\bibfnamefont {P.}~\bibnamefont {Schmitteckert}},\
  }\bibfield  {title} {\enquote {\bibinfo {title} {Twofold advance in the
  theoretical understanding of far-from-equilibrium properties of interacting
  nanostructures},}\ }\href {\doibase 10.1103/PhysRevLett.101.140601}
  {\bibfield  {journal} {\bibinfo  {journal} {Phys. Rev. Lett.}\ }\textbf
  {\bibinfo {volume} {101}},\ \bibinfo {pages} {140601} (\bibinfo {year}
  {2008})}\BibitemShut {NoStop}%
\bibitem [{\citenamefont {Ralph}\ and\ \citenamefont
  {Buhrman}(1994)}]{Ralph94}%
  \BibitemOpen
  \bibfield  {author} {\bibinfo {author} {\bibfnamefont {D.~C.}\ \bibnamefont
  {Ralph}}\ and\ \bibinfo {author} {\bibfnamefont {R.~A.}\ \bibnamefont
  {Buhrman}},\ }\bibfield  {title} {\enquote {\bibinfo {title}
  {{Kondo}-assisted and resonant tunneling via a single charge trap: A
  realization of the {Anderson} model out of equilibrium},}\ }\href {\doibase
  10.1103/PhysRevLett.72.3401} {\bibfield  {journal} {\bibinfo  {journal}
  {Phys. Rev. Lett.}\ }\textbf {\bibinfo {volume} {72}},\ \bibinfo {pages}
  {3401--3404} (\bibinfo {year} {1994})}\BibitemShut {NoStop}%
\bibitem [{\citenamefont {Goldhaber-Gordon}\ \emph {et~al.}(1998)\citenamefont
  {Goldhaber-Gordon}, \citenamefont {Shtrikman}, \citenamefont {Mahalu},
  \citenamefont {Abusch-Magder}, \citenamefont {Meirav},\ and\ \citenamefont
  {Kastner}}]{Goldhaber98}%
  \BibitemOpen
  \bibfield  {author} {\bibinfo {author} {\bibfnamefont {D.}~\bibnamefont
  {Goldhaber-Gordon}}, \bibinfo {author} {\bibfnamefont {H.}~\bibnamefont
  {Shtrikman}}, \bibinfo {author} {\bibfnamefont {D.}~\bibnamefont {Mahalu}},
  \bibinfo {author} {\bibfnamefont {D.}~\bibnamefont {Abusch-Magder}}, \bibinfo
  {author} {\bibfnamefont {U.}~\bibnamefont {Meirav}}, \ and\ \bibinfo {author}
  {\bibfnamefont {M.~A.}\ \bibnamefont {Kastner}},\ }\bibfield  {title}
  {\enquote {\bibinfo {title} {{Kondo} effect in a single-electron
  transistor},}\ }\href {http://dx.doi.org/10.1038/34373} {\bibfield  {journal}
  {\bibinfo  {journal} {Nature}\ }\textbf {\bibinfo {volume} {391}},\ \bibinfo
  {pages} {156--159} (\bibinfo {year} {1998})}\BibitemShut {NoStop}%
\bibitem [{\citenamefont {Cronenwett}\ \emph {et~al.}(1998)\citenamefont
  {Cronenwett}, \citenamefont {Oosterkamp},\ and\ \citenamefont
  {Kouwenhoven}}]{Cronenwett98}%
  \BibitemOpen
  \bibfield  {author} {\bibinfo {author} {\bibfnamefont {S.~M.}\ \bibnamefont
  {Cronenwett}}, \bibinfo {author} {\bibfnamefont {T.~H.}\ \bibnamefont
  {Oosterkamp}}, \ and\ \bibinfo {author} {\bibfnamefont {L.~P.}\ \bibnamefont
  {Kouwenhoven}},\ }\bibfield  {title} {\enquote {\bibinfo {title} {A tunable
  {Kondo} effect in quantum dots},}\ }\href {\doibase
  10.1126/science.281.5376.540} {\bibfield  {journal} {\bibinfo  {journal}
  {Science}\ }\textbf {\bibinfo {volume} {281}},\ \bibinfo {pages} {540--544}
  (\bibinfo {year} {1998})}\BibitemShut {NoStop}%
\bibitem [{\citenamefont {Simmel}\ \emph {et~al.}(1999)\citenamefont {Simmel},
  \citenamefont {Blick}, \citenamefont {Kotthaus}, \citenamefont
  {Wegscheider},\ and\ \citenamefont {Bichler}}]{Simmel99}%
  \BibitemOpen
  \bibfield  {author} {\bibinfo {author} {\bibfnamefont {F.}~\bibnamefont
  {Simmel}}, \bibinfo {author} {\bibfnamefont {R.~H.}\ \bibnamefont {Blick}},
  \bibinfo {author} {\bibfnamefont {J.~P.}\ \bibnamefont {Kotthaus}}, \bibinfo
  {author} {\bibfnamefont {W.}~\bibnamefont {Wegscheider}}, \ and\ \bibinfo
  {author} {\bibfnamefont {M.}~\bibnamefont {Bichler}},\ }\bibfield  {title}
  {\enquote {\bibinfo {title} {Anomalous {Kondo} effect in a quantum dot at
  nonzero bias},}\ }\href {\doibase 10.1103/PhysRevLett.83.804} {\bibfield
  {journal} {\bibinfo  {journal} {Phys. Rev. Lett.}\ }\textbf {\bibinfo
  {volume} {83}},\ \bibinfo {pages} {804--807} (\bibinfo {year}
  {1999})}\BibitemShut {NoStop}%
\bibitem [{\citenamefont {van~der Wiel}\ \emph {et~al.}(2000)\citenamefont
  {van~der Wiel}, \citenamefont {Franceschi}, \citenamefont {Fujisawa},
  \citenamefont {Elzerman}, \citenamefont {Tarucha},\ and\ \citenamefont
  {Kouwenhoven}}]{vanderWiel00}%
  \BibitemOpen
  \bibfield  {author} {\bibinfo {author} {\bibfnamefont {W.~G.}\ \bibnamefont
  {van~der Wiel}}, \bibinfo {author} {\bibfnamefont {S.~De}\ \bibnamefont
  {Franceschi}}, \bibinfo {author} {\bibfnamefont {T.}~\bibnamefont
  {Fujisawa}}, \bibinfo {author} {\bibfnamefont {J.~M.}\ \bibnamefont
  {Elzerman}}, \bibinfo {author} {\bibfnamefont {S.}~\bibnamefont {Tarucha}}, \
  and\ \bibinfo {author} {\bibfnamefont {L.~P.}\ \bibnamefont {Kouwenhoven}},\
  }\bibfield  {title} {\enquote {\bibinfo {title} {The {Kondo} effect in the
  unitary limit},}\ }\href {\doibase 10.1126/science.289.5487.2105} {\bibfield
  {journal} {\bibinfo  {journal} {Science}\ }\textbf {\bibinfo {volume}
  {289}},\ \bibinfo {pages} {2105--2108} (\bibinfo {year} {2000})}\BibitemShut
  {NoStop}%
\bibitem [{\citenamefont {Kretinin}\ \emph {et~al.}(2011)\citenamefont
  {Kretinin}, \citenamefont {Shtrikman}, \citenamefont {Goldhaber-Gordon},
  \citenamefont {Hanl}, \citenamefont {Weichselbaum}, \citenamefont {von
  Delft}, \citenamefont {Costi},\ and\ \citenamefont {Mahalu}}]{Kretinin11}%
  \BibitemOpen
  \bibfield  {author} {\bibinfo {author} {\bibfnamefont {A.~V.}\ \bibnamefont
  {Kretinin}}, \bibinfo {author} {\bibfnamefont {H.}~\bibnamefont {Shtrikman}},
  \bibinfo {author} {\bibfnamefont {D.}~\bibnamefont {Goldhaber-Gordon}},
  \bibinfo {author} {\bibfnamefont {M.}~\bibnamefont {Hanl}}, \bibinfo {author}
  {\bibfnamefont {A.}~\bibnamefont {Weichselbaum}}, \bibinfo {author}
  {\bibfnamefont {J.}~\bibnamefont {von Delft}}, \bibinfo {author}
  {\bibfnamefont {T.}~\bibnamefont {Costi}}, \ and\ \bibinfo {author}
  {\bibfnamefont {D.}~\bibnamefont {Mahalu}},\ }\bibfield  {title} {\enquote
  {\bibinfo {title} {Spin-$\frac{1}{2}$ {Kondo} effect in an {InAs} nanowire
  quantum dot: Unitary limit, conductance scaling, and {Zeeman} splitting},}\
  }\href {\doibase 10.1103/PhysRevB.84.245316} {\bibfield  {journal} {\bibinfo
  {journal} {Phys. Rev. B}\ }\textbf {\bibinfo {volume} {84}},\ \bibinfo
  {pages} {245316} (\bibinfo {year} {2011})}\BibitemShut {NoStop}%
\bibitem [{\citenamefont {Kretinin}\ \emph {et~al.}(2012)\citenamefont
  {Kretinin}, \citenamefont {Shtrikman},\ and\ \citenamefont
  {Mahalu}}]{Kretinin12}%
  \BibitemOpen
  \bibfield  {author} {\bibinfo {author} {\bibfnamefont {A.~V.}\ \bibnamefont
  {Kretinin}}, \bibinfo {author} {\bibfnamefont {H.}~\bibnamefont {Shtrikman}},
  \ and\ \bibinfo {author} {\bibfnamefont {D.}~\bibnamefont {Mahalu}},\
  }\bibfield  {title} {\enquote {\bibinfo {title} {Universal line shape of the
  {Kondo} zero-bias anomaly in a quantum dot},}\ }\href {\doibase
  10.1103/PhysRevB.85.201301} {\bibfield  {journal} {\bibinfo  {journal} {Phys.
  Rev. B}\ }\textbf {\bibinfo {volume} {85}},\ \bibinfo {pages} {201301}
  (\bibinfo {year} {2012})}\BibitemShut {NoStop}%
\bibitem [{\citenamefont {Wilson}(1975)}]{Wilson75}%
  \BibitemOpen
  \bibfield  {author} {\bibinfo {author} {\bibfnamefont {K.~G.}\ \bibnamefont
  {Wilson}},\ }\bibfield  {title} {\enquote {\bibinfo {title} {The
  renormalization group: {Critical} phenomena and the {Kondo} problem},}\
  }\href {\doibase 10.1103/RevModPhys.47.773} {\bibfield  {journal} {\bibinfo
  {journal} {Rev. Mod. Phys.}\ }\textbf {\bibinfo {volume} {47}},\ \bibinfo
  {pages} {773--840} (\bibinfo {year} {1975})}\BibitemShut {NoStop}%
\bibitem [{\citenamefont {Bulla}\ \emph {et~al.}(2008)\citenamefont {Bulla},
  \citenamefont {Costi},\ and\ \citenamefont {Pruschke}}]{Bulla08}%
  \BibitemOpen
  \bibfield  {author} {\bibinfo {author} {\bibfnamefont {R.}~\bibnamefont
  {Bulla}}, \bibinfo {author} {\bibfnamefont {T.~A.}\ \bibnamefont {Costi}}, \
  and\ \bibinfo {author} {\bibfnamefont {T.}~\bibnamefont {Pruschke}},\
  }\bibfield  {title} {\enquote {\bibinfo {title} {Numerical renormalization
  group method for quantum impurity systems},}\ }\href {\doibase
  10.1103/RevModPhys.80.395} {\bibfield  {journal} {\bibinfo  {journal} {Rev.
  Mod. Phys.}\ }\textbf {\bibinfo {volume} {80}},\ \bibinfo {pages} {395--450}
  (\bibinfo {year} {2008})}\BibitemShut {NoStop}%
\bibitem [{\citenamefont {G\"uttge}\ \emph {et~al.}(2013)\citenamefont
  {G\"uttge}, \citenamefont {Anders}, \citenamefont {Schollw\"ock},
  \citenamefont {Eidelstein},\ and\ \citenamefont {Schiller}}]{Guttge2013}%
  \BibitemOpen
  \bibfield  {author} {\bibinfo {author} {\bibfnamefont {F.}~\bibnamefont
  {G\"uttge}}, \bibinfo {author} {\bibfnamefont {F.~B.}\ \bibnamefont
  {Anders}}, \bibinfo {author} {\bibfnamefont {U.}~\bibnamefont
  {Schollw\"ock}}, \bibinfo {author} {\bibfnamefont {E.}~\bibnamefont
  {Eidelstein}}, \ and\ \bibinfo {author} {\bibfnamefont {A.}~\bibnamefont
  {Schiller}},\ }\bibfield  {title} {\enquote {\bibinfo {title} {Hybrid
  {NRG}-{DMRG} approach to real-time dynamics of quantum impurity systems},}\
  }\href {\doibase 10.1103/PhysRevB.87.115115} {\bibfield  {journal} {\bibinfo
  {journal} {Phys. Rev. B}\ }\textbf {\bibinfo {volume} {87}},\ \bibinfo
  {pages} {115115} (\bibinfo {year} {2013})}\BibitemShut {NoStop}%
\bibitem [{\citenamefont {Vidal}(2004)}]{Vidal04}%
  \BibitemOpen
  \bibfield  {author} {\bibinfo {author} {\bibfnamefont {G.}~\bibnamefont
  {Vidal}},\ }\bibfield  {title} {\enquote {\bibinfo {title} {Efficient
  simulation of one-dimensional quantum many-body systems},}\ }\href {\doibase
  10.1103/PhysRevLett.93.040502} {\bibfield  {journal} {\bibinfo  {journal}
  {Phys. Rev. Lett.}\ }\textbf {\bibinfo {volume} {93}},\ \bibinfo {pages}
  {040502} (\bibinfo {year} {2004})}\BibitemShut {NoStop}%
\bibitem [{\citenamefont {Daley}\ \emph {et~al.}(2004)\citenamefont {Daley},
  \citenamefont {Kollath}, \citenamefont {Schollw\"ock},\ and\ \citenamefont
  {Vidal}}]{Daley04}%
  \BibitemOpen
  \bibfield  {author} {\bibinfo {author} {\bibfnamefont {A.~J.}\ \bibnamefont
  {Daley}}, \bibinfo {author} {\bibfnamefont {C.}~\bibnamefont {Kollath}},
  \bibinfo {author} {\bibfnamefont {U.}~\bibnamefont {Schollw\"ock}}, \ and\
  \bibinfo {author} {\bibfnamefont {G.}~\bibnamefont {Vidal}},\ }\bibfield
  {title} {\enquote {\bibinfo {title} {Time-dependent density-matrix
  renormalization-group using adaptive effective {Hilbert} spaces},}\ }\href
  {http://stacks.iop.org/1742-5468/2004/i=04/a=P04005} {\bibfield  {journal}
  {\bibinfo  {journal} {J. Stat. Mech.}\ }\textbf {\bibinfo {volume}
  {(2004)}},\ \bibinfo {pages} {P04005} (\bibinfo {year} {2004})}\BibitemShut
  {NoStop}%
\bibitem [{\citenamefont {White}\ and\ \citenamefont
  {Feiguin}(2004)}]{White04}%
  \BibitemOpen
  \bibfield  {author} {\bibinfo {author} {\bibfnamefont {S.~R.}\ \bibnamefont
  {White}}\ and\ \bibinfo {author} {\bibfnamefont {A.~E.}\ \bibnamefont
  {Feiguin}},\ }\bibfield  {title} {\enquote {\bibinfo {title} {Real-time
  evolution using the density matrix renormalization group},}\ }\href {\doibase
  10.1103/PhysRevLett.93.076401} {\bibfield  {journal} {\bibinfo  {journal}
  {Phys. Rev. Lett.}\ }\textbf {\bibinfo {volume} {93}},\ \bibinfo {pages}
  {076401} (\bibinfo {year} {2004})}\BibitemShut {NoStop}%
\bibitem [{\citenamefont {Schollw\"ock}(2011)}]{Schollwoeck11}%
  \BibitemOpen
  \bibfield  {author} {\bibinfo {author} {\bibfnamefont {U.}~\bibnamefont
  {Schollw\"ock}},\ }\bibfield  {title} {\enquote {\bibinfo {title} {The
  density-matrix renormalization group in the age of matrix product states},}\
  }\href {\doibase 10.1016/j.aop.2010.09.012} {\bibfield  {journal} {\bibinfo
  {journal} {Ann.\ Phys.}\ }\textbf {\bibinfo {volume} {326}},\ \bibinfo
  {pages} {96 -- 192} (\bibinfo {year} {2011})}\BibitemShut {NoStop}%
\bibitem [{\citenamefont {Bransch\"adel}\ \emph {et~al.}(2010)\citenamefont
  {Bransch\"adel}, \citenamefont {Schneider},\ and\ \citenamefont
  {Schmitteckert}}]{Branschadel10}%
  \BibitemOpen
  \bibfield  {author} {\bibinfo {author} {\bibfnamefont {A.}~\bibnamefont
  {Bransch\"adel}}, \bibinfo {author} {\bibfnamefont {G.}~\bibnamefont
  {Schneider}}, \ and\ \bibinfo {author} {\bibfnamefont {P.}~\bibnamefont
  {Schmitteckert}},\ }\bibfield  {title} {\enquote {\bibinfo {title}
  {Conductance of inhomogeneous systems: Real-time dynamics},}\ }\href
  {\doibase 10.1002/andp.201000017} {\bibfield  {journal} {\bibinfo  {journal}
  {Ann.\ Phys.}\ }\textbf {\bibinfo {volume} {522}},\ \bibinfo {pages}
  {657--678} (\bibinfo {year} {2010})}\BibitemShut {NoStop}%
\bibitem [{\citenamefont {Dias~da Silva}\ \emph {et~al.}(2008)\citenamefont
  {Dias~da Silva}, \citenamefont {Heidrich-Meisner}, \citenamefont {Feiguin},
  \citenamefont {B\"usser}, \citenamefont {Martins}, \citenamefont {Anda},\
  and\ \citenamefont {Dagotto}}]{daSilva08}%
  \BibitemOpen
  \bibfield  {author} {\bibinfo {author} {\bibfnamefont {L.~G. G.~V.}\
  \bibnamefont {Dias~da Silva}}, \bibinfo {author} {\bibfnamefont
  {F.}~\bibnamefont {Heidrich-Meisner}}, \bibinfo {author} {\bibfnamefont
  {A.~E.}\ \bibnamefont {Feiguin}}, \bibinfo {author} {\bibfnamefont {C.~A.}\
  \bibnamefont {B\"usser}}, \bibinfo {author} {\bibfnamefont {G.~B.}\
  \bibnamefont {Martins}}, \bibinfo {author} {\bibfnamefont {E.~V.}\
  \bibnamefont {Anda}}, \ and\ \bibinfo {author} {\bibfnamefont
  {E.}~\bibnamefont {Dagotto}},\ }\bibfield  {title} {\enquote {\bibinfo
  {title} {Transport properties and {Kondo} correlations in nanostructures:
  Time-dependent {DMRG} method applied to quantum dots coupled to {Wilson}
  chains},}\ }\href {\doibase 10.1103/PhysRevB.78.195317} {\bibfield  {journal}
  {\bibinfo  {journal} {Phys. Rev. B}\ }\textbf {\bibinfo {volume} {78}},\
  \bibinfo {pages} {195317} (\bibinfo {year} {2008})}\BibitemShut {NoStop}%
\bibitem [{\citenamefont {Takahashi}\ and\ \citenamefont
  {Umezawa}(1975)}]{Takahashi75}%
  \BibitemOpen
  \bibfield  {author} {\bibinfo {author} {\bibfnamefont {Y.}~\bibnamefont
  {Takahashi}}\ and\ \bibinfo {author} {\bibfnamefont {H.}~\bibnamefont
  {Umezawa}},\ }\bibfield  {title} {\enquote {\bibinfo {title} {Thermo field
  dynamics},}\ }\href@noop {} {\bibfield  {journal} {\bibinfo  {journal}
  {Collective Phenomena}\ }\textbf {\bibinfo {volume} {2}},\ \bibinfo {pages}
  {55–80} (\bibinfo {year} {1975})}\BibitemShut {NoStop}%
\bibitem [{\citenamefont {Barnett}\ and\ \citenamefont
  {Dalton}(1987)}]{Barnett87}%
  \BibitemOpen
  \bibfield  {author} {\bibinfo {author} {\bibfnamefont {S.~M.}\ \bibnamefont
  {Barnett}}\ and\ \bibinfo {author} {\bibfnamefont {B.~J.}\ \bibnamefont
  {Dalton}},\ }\bibfield  {title} {\enquote {\bibinfo {title} {Liouville space
  description of thermofields and their generalisations},}\ }\href
  {http://stacks.iop.org/0305-4470/20/i=2/a=026} {\bibfield  {journal}
  {\bibinfo  {journal} {Journal of Physics A: Mathematical and General}\
  }\textbf {\bibinfo {volume} {20}},\ \bibinfo {pages} {411} (\bibinfo {year}
  {1987})}\BibitemShut {NoStop}%
\bibitem [{\citenamefont {Das}(2000)}]{Das00}%
  \BibitemOpen
  \bibfield  {author} {\bibinfo {author} {\bibfnamefont {A.}~\bibnamefont
  {Das}},\ }\enquote {\bibinfo {title} {Topics in finite temperature field
  theory},}\ in\ \href {https://arxiv.org/abs/hep-ph/0004125} {\emph {\bibinfo
  {booktitle} {Quantum Field Theory - A 20th Century Profile}}},\ \bibinfo
  {editor} {edited by\ \bibinfo {editor} {\bibfnamefont {Asoke~N.}\
  \bibnamefont {Mitra}}}\ (\bibinfo  {publisher} {Hindustan Book Agency},\
  \bibinfo {address} {New Delhi},\ \bibinfo {year} {2000})\ pp.\ \bibinfo
  {pages} {383--411}\BibitemShut {NoStop}%
\bibitem [{\citenamefont {de~Vega}\ and\ \citenamefont
  {Ba\~nuls}(2015)}]{deVega15}%
  \BibitemOpen
  \bibfield  {author} {\bibinfo {author} {\bibfnamefont {I.}~\bibnamefont
  {de~Vega}}\ and\ \bibinfo {author} {\bibfnamefont {M.-C.}\ \bibnamefont
  {Ba\~nuls}},\ }\bibfield  {title} {\enquote {\bibinfo {title}
  {Thermofield-based chain-mapping approach for open quantum systems},}\ }\href
  {\doibase 10.1103/PhysRevA.92.052116} {\bibfield  {journal} {\bibinfo
  {journal} {Phys. Rev. A}\ }\textbf {\bibinfo {volume} {92}},\ \bibinfo
  {pages} {052116} (\bibinfo {year} {2015})}\BibitemShut {NoStop}%
\bibitem [{\citenamefont {Guo}\ \emph {et~al.}(2018)\citenamefont {Guo},
  \citenamefont {de~Vega}, \citenamefont {Schollw\"ock},\ and\ \citenamefont
  {Poletti}}]{Guo17}%
  \BibitemOpen
  \bibfield  {author} {\bibinfo {author} {\bibfnamefont {C.}~\bibnamefont
  {Guo}}, \bibinfo {author} {\bibfnamefont {I.}~\bibnamefont {de~Vega}},
  \bibinfo {author} {\bibfnamefont {U.}~\bibnamefont {Schollw\"ock}}, \ and\
  \bibinfo {author} {\bibfnamefont {D.}~\bibnamefont {Poletti}},\ }\bibfield
  {title} {\enquote {\bibinfo {title} {Stable-unstable transition for a
  {Bose-Hubbard} chain coupled to an environment},}\ }\href
  {10.1103/PhysRevA.97.053610} {\bibfield  {journal} {\bibinfo  {journal}
  {Phys. Rev. A}\ }\textbf {\bibinfo {volume} {97}},\ \bibinfo {pages} {053610}
  (\bibinfo {year} {2018})},\ \bibinfo {note} {arXiv:1708.01939}\BibitemShut
  {NoStop}%
\bibitem [{sup()}]{sup}%
  \BibitemOpen
  \href@noop {} {\bibinfo  {journal} {See Supplementary Material at [URL will
  be inserted by publisher], which includes
  Refs.~\cite{Schmitteckert10,Campo05,Zitko09,Wb11_rho,*Wb12_FDM,*Wb12_SUN,Corboz10,Schneider06,Wang10,Barthel09,Hanl14},
  for details}\ }\BibitemShut {NoStop}%
\bibitem [{\citenamefont {Campo}\ and\ \citenamefont
  {Oliveira}(2005)}]{Campo05}%
  \BibitemOpen
\bibfield  {journal} {  }\bibfield  {author} {\bibinfo {author} {\bibfnamefont
  {V.~L.}\ \bibnamefont {Campo}}\ and\ \bibinfo {author} {\bibfnamefont
  {L.~N.}\ \bibnamefont {Oliveira}},\ }\bibfield  {title} {\enquote {\bibinfo
  {title} {Alternative discretization in the numerical renormalization-group
  method},}\ }\href {\doibase 10.1103/PhysRevB.72.104432} {\bibfield  {journal}
  {\bibinfo  {journal} {Phys. Rev. B}\ }\textbf {\bibinfo {volume} {72}},\
  \bibinfo {pages} {104432} (\bibinfo {year} {2005})}\BibitemShut {NoStop}%
\bibitem [{\citenamefont {\v{Z}itko}(2009)}]{Zitko09}%
  \BibitemOpen
  \bibfield  {author} {\bibinfo {author} {\bibfnamefont {R.}~\bibnamefont
  {\v{Z}itko}},\ }\bibfield  {title} {\enquote {\bibinfo {title} {Adaptive
  logarithmic discretization for numerical renormalization group methods},}\
  }\href {\doibase 10.1016/j.cpc.2009.02.007} {\bibfield  {journal} {\bibinfo
  {journal} {Comput. Phys. Commun.}\ }\textbf {\bibinfo {volume} {180}},\
  \bibinfo {pages} {1271 -- 1276} (\bibinfo {year} {2009})}\BibitemShut
  {NoStop}%
\bibitem [{\citenamefont {Weichselbaum}(2011)}]{Wb11_rho}%
  \BibitemOpen
  \bibfield  {author} {\bibinfo {author} {\bibfnamefont {A.}~\bibnamefont
  {Weichselbaum}},\ }\bibfield  {title} {\enquote {\bibinfo {title} {Discarded
  weight and entanglement spectra in the numerical renormalization group},}\
  }\href {\doibase 10.1103/PhysRevB.84.125130} {\bibfield  {journal} {\bibinfo
  {journal} {Phys. Rev. B}\ }\textbf {\bibinfo {volume} {84}},\ \bibinfo
  {pages} {125130} (\bibinfo {year} {2011})}\BibitemShut {NoStop}%
\bibitem [{\citenamefont {Weichselbaum}(2012{\natexlab{a}})}]{Wb12_FDM}%
  \BibitemOpen
  \bibfield  {author} {\bibinfo {author} {\bibfnamefont {A.}~\bibnamefont
  {Weichselbaum}},\ }\bibfield  {title} {\enquote {\bibinfo {title} {Tensor
  networks and the numerical renormalization group},}\ }\href {\doibase
  10.1103/PhysRevB.86.245124} {\bibfield  {journal} {\bibinfo  {journal} {Phys.
  Rev. B}\ }\textbf {\bibinfo {volume} {86}},\ \bibinfo {pages} {245124}
  (\bibinfo {year} {2012}{\natexlab{a}})}\BibitemShut {NoStop}%
\bibitem [{\citenamefont {Weichselbaum}(2012{\natexlab{b}})}]{Wb12_SUN}%
  \BibitemOpen
  \bibfield  {author} {\bibinfo {author} {\bibfnamefont {A.}~\bibnamefont
  {Weichselbaum}},\ }\bibfield  {title} {\enquote {\bibinfo {title}
  {{Non-Abelian} symmetries in tensor networks: A quantum symmetry space
  approach},}\ }\href {\doibase 10.1016/j.aop.2012.07.009} {\bibfield
  {journal} {\bibinfo  {journal} {Ann. of Phys.}\ }\textbf {\bibinfo {volume}
  {327}},\ \bibinfo {pages} {2972 -- 3047} (\bibinfo {year}
  {2012}{\natexlab{b}})}\BibitemShut {NoStop}%
\bibitem [{\citenamefont {Corboz}\ \emph {et~al.}(2010)\citenamefont {Corboz},
  \citenamefont {Or{\'u}s}, \citenamefont {Bauer},\ and\ \citenamefont
  {Vidal}}]{Corboz10}%
  \BibitemOpen
  \bibfield  {author} {\bibinfo {author} {\bibfnamefont {P.}~\bibnamefont
  {Corboz}}, \bibinfo {author} {\bibfnamefont {R.}~\bibnamefont {Or{\'u}s}},
  \bibinfo {author} {\bibfnamefont {B.}~\bibnamefont {Bauer}}, \ and\ \bibinfo
  {author} {\bibfnamefont {G.}~\bibnamefont {Vidal}},\ }\bibfield  {title}
  {\enquote {\bibinfo {title} {Simulation of strongly correlated fermions in
  two spatial dimensions with fermionic projected entangled-pair states},}\
  }\href {\doibase 10.1103/PhysRevB.81.165104} {\bibfield  {journal} {\bibinfo
  {journal} {Phys. Rev. B}\ }\textbf {\bibinfo {volume} {81}},\ \bibinfo
  {pages} {165104} (\bibinfo {year} {2010})}\BibitemShut {NoStop}%
\bibitem [{\citenamefont {Schneider}\ and\ \citenamefont
  {Schmitteckert}(2006)}]{Schneider06}%
  \BibitemOpen
  \bibfield  {author} {\bibinfo {author} {\bibfnamefont {G.}~\bibnamefont
  {Schneider}}\ and\ \bibinfo {author} {\bibfnamefont {P.}~\bibnamefont
  {Schmitteckert}},\ }\bibfield  {title} {\enquote {\bibinfo {title}
  {Conductance in strongly correlated {1D} systems: Real-time dynamics in
  {DMRG}},}\ }\href {https://arxiv.org/abs/cond-mat/0601389} {\bibfield
  {journal} {\bibinfo  {journal} {arXiv:cond-mat/0601389}\ } (\bibinfo {year}
  {2006})}\BibitemShut {NoStop}%
\bibitem [{\citenamefont {Wang}\ and\ \citenamefont {Kehrein}(2010)}]{Wang10}%
  \BibitemOpen
  \bibfield  {author} {\bibinfo {author} {\bibfnamefont {P.}~\bibnamefont
  {Wang}}\ and\ \bibinfo {author} {\bibfnamefont {S.}~\bibnamefont {Kehrein}},\
  }\bibfield  {title} {\enquote {\bibinfo {title} {Flow equation calculation of
  transient and steady-state currents in the {Anderson} impurity model},}\
  }\href {\doibase 10.1103/PhysRevB.82.125124} {\bibfield  {journal} {\bibinfo
  {journal} {Phys. Rev. B}\ }\textbf {\bibinfo {volume} {82}},\ \bibinfo
  {pages} {125124} (\bibinfo {year} {2010})}\BibitemShut {NoStop}%
\bibitem [{\citenamefont {Barthel}\ \emph {et~al.}(2009)\citenamefont
  {Barthel}, \citenamefont {Schollw\"ock},\ and\ \citenamefont
  {White}}]{Barthel09}%
  \BibitemOpen
  \bibfield  {author} {\bibinfo {author} {\bibfnamefont {T.}~\bibnamefont
  {Barthel}}, \bibinfo {author} {\bibfnamefont {U.}~\bibnamefont
  {Schollw\"ock}}, \ and\ \bibinfo {author} {\bibfnamefont {S.~R.}\
  \bibnamefont {White}},\ }\bibfield  {title} {\enquote {\bibinfo {title}
  {Spectral functions in one-dimensional quantum systems at finite temperature
  using the density matrix renormalization group},}\ }\href {\doibase
  10.1103/PhysRevB.79.245101} {\bibfield  {journal} {\bibinfo  {journal} {Phys.
  Rev. B}\ }\textbf {\bibinfo {volume} {79}},\ \bibinfo {pages} {245101}
  (\bibinfo {year} {2009})}\BibitemShut {NoStop}%
\bibitem [{\citenamefont {Schmitteckert}(2010)}]{Schmitteckert10}%
  \BibitemOpen
  \bibfield  {author} {\bibinfo {author} {\bibfnamefont {P.}~\bibnamefont
  {Schmitteckert}},\ }\bibfield  {title} {\enquote {\bibinfo {title}
  {Calculating {Green} functions from finite systems},}\ }\href
  {http://stacks.iop.org/1742-6596/220/i=1/a=012022} {\bibfield  {journal}
  {\bibinfo  {journal} {J.\ Phys.:\ Conf.\ Series}\ }\textbf {\bibinfo {volume}
  {220}},\ \bibinfo {pages} {012022} (\bibinfo {year} {2010})}\BibitemShut
  {NoStop}%
\bibitem [{\citenamefont {Hanl}\ and\ \citenamefont
  {Weichselbaum}(2014)}]{Hanl14}%
  \BibitemOpen
  \bibfield  {author} {\bibinfo {author} {\bibfnamefont {M.}~\bibnamefont
  {Hanl}}\ and\ \bibinfo {author} {\bibfnamefont {A.}~\bibnamefont
  {Weichselbaum}},\ }\bibfield  {title} {\enquote {\bibinfo {title} {Local
  susceptibility and {Kondo} scaling in the presence of finite bandwidth},}\
  }\href {\doibase 10.1103/PhysRevB.89.075130} {\bibfield  {journal} {\bibinfo
  {journal} {Phys. Rev. B}\ }\textbf {\bibinfo {volume} {89}},\ \bibinfo
  {pages} {075130} (\bibinfo {year} {2014})}\BibitemShut {NoStop}%
\bibitem [{\citenamefont {Karrasch}\ \emph {et~al.}(2012)\citenamefont
  {Karrasch}, \citenamefont {Bardarson},\ and\ \citenamefont
  {Moore}}]{Karrasch12}%
  \BibitemOpen
  \bibfield  {author} {\bibinfo {author} {\bibfnamefont {C.}~\bibnamefont
  {Karrasch}}, \bibinfo {author} {\bibfnamefont {J.~H.}\ \bibnamefont
  {Bardarson}}, \ and\ \bibinfo {author} {\bibfnamefont {J.~E.}\ \bibnamefont
  {Moore}},\ }\bibfield  {title} {\enquote {\bibinfo {title}
  {Finite-temperature dynamical density matrix renormalization group and the
  {Drude} weight of spin-$1/2$ chains},}\ }\href {\doibase
  10.1103/PhysRevLett.108.227206} {\bibfield  {journal} {\bibinfo  {journal}
  {Phys. Rev. Lett.}\ }\textbf {\bibinfo {volume} {108}},\ \bibinfo {pages}
  {227206} (\bibinfo {year} {2012})}\BibitemShut {NoStop}%
\bibitem [{\citenamefont {Bidzhiev}\ and\ \citenamefont
  {Misguich}(2017)}]{Bidzhiev17}%
  \BibitemOpen
  \bibfield  {author} {\bibinfo {author} {\bibfnamefont {K.}~\bibnamefont
  {Bidzhiev}}\ and\ \bibinfo {author} {\bibfnamefont {G.}~\bibnamefont
  {Misguich}},\ }\bibfield  {title} {\enquote {\bibinfo {title}
  {Out-of-equilibrium dynamics in a quantum impurity model: Numerics for
  particle transport and entanglement entropy},}\ }\href {\doibase
  10.1103/PhysRevB.96.195117} {\bibfield  {journal} {\bibinfo  {journal} {Phys.
  Rev. B}\ }\textbf {\bibinfo {volume} {96}},\ \bibinfo {pages} {195117}
  (\bibinfo {year} {2017})}\BibitemShut {NoStop}%
\bibitem [{\citenamefont {Wiegmann}\ and\ \citenamefont
  {Tsvelick}(1983)}]{Wiegmann83}%
  \BibitemOpen
  \bibfield  {author} {\bibinfo {author} {\bibfnamefont {P.~B.}\ \bibnamefont
  {Wiegmann}}\ and\ \bibinfo {author} {\bibfnamefont {A.~M.}\ \bibnamefont
  {Tsvelick}},\ }\bibfield  {title} {\enquote {\bibinfo {title} {Exact solution
  of the {Anderson} model: {I}},}\ }\href
  {http://stacks.iop.org/0022-3719/16/i=12/a=017} {\bibfield  {journal}
  {\bibinfo  {journal} {J. of Phys. C: Solid State Phys.}\ }\textbf {\bibinfo
  {volume} {16}},\ \bibinfo {pages} {2281} (\bibinfo {year}
  {1983})}\BibitemShut {NoStop}%
\bibitem [{\citenamefont {Tsvelick}\ and\ \citenamefont
  {Wiegmann}(1983)}]{Tsvelick83}%
  \BibitemOpen
  \bibfield  {author} {\bibinfo {author} {\bibfnamefont {A.~M.}\ \bibnamefont
  {Tsvelick}}\ and\ \bibinfo {author} {\bibfnamefont {P.~B.}\ \bibnamefont
  {Wiegmann}},\ }\bibfield  {title} {\enquote {\bibinfo {title} {Exact solution
  of the {Anderson} model. {II. Thermodynamic} properties at finite
  temperatures},}\ }\href {http://stacks.iop.org/0022-3719/16/i=12/a=018}
  {\bibfield  {journal} {\bibinfo  {journal} {J. Phys. C: Solid State Phys.}\
  }\textbf {\bibinfo {volume} {16}},\ \bibinfo {pages} {2321} (\bibinfo {year}
  {1983})}\BibitemShut {NoStop}%
\bibitem [{\citenamefont {Fugger}\ \emph {et~al.}(2018)\citenamefont {Fugger},
  \citenamefont {Dorda}, \citenamefont {Schwarz}, \citenamefont {von Delft},\
  and\ \citenamefont {Arrigoni}}]{Fugger17}%
  \BibitemOpen
  \bibfield  {author} {\bibinfo {author} {\bibfnamefont {D.~M.}\ \bibnamefont
  {Fugger}}, \bibinfo {author} {\bibfnamefont {A.}~\bibnamefont {Dorda}},
  \bibinfo {author} {\bibfnamefont {F.}~\bibnamefont {Schwarz}}, \bibinfo
  {author} {\bibfnamefont {J.}~\bibnamefont {von Delft}}, \ and\ \bibinfo
  {author} {\bibfnamefont {E.}~\bibnamefont {Arrigoni}},\ }\bibfield  {title}
  {\enquote {\bibinfo {title} {Nonequilibrium {Kondo} effect in a magnetic
  field: auxiliary master equation approach},}\ }\href {\doibase
  10.1088/1367-2630/aa9fdc/meta} {\bibfield  {journal} {\bibinfo  {journal}
  {New. J. Phys.}\ }\textbf {\bibinfo {volume} {20}},\ \bibinfo {pages}
  {013030} (\bibinfo {year} {2018})}\BibitemShut {NoStop}%
\bibitem [{\citenamefont {Weichselbaum}\ and\ \citenamefont {von
  Delft}(2007)}]{AW07}%
  \BibitemOpen
  \bibfield  {author} {\bibinfo {author} {\bibfnamefont {A.}~\bibnamefont
  {Weichselbaum}}\ and\ \bibinfo {author} {\bibfnamefont {J.}~\bibnamefont {von
  Delft}},\ }\bibfield  {title} {\enquote {\bibinfo {title} {Sum-rule
  conserving spectral functions from the numerical renormalization group},}\
  }\href {\doibase 10.1103/PhysRevLett.99.076402} {\bibfield  {journal}
  {\bibinfo  {journal} {Phys. Rev. Lett.}\ }\textbf {\bibinfo {volume} {99}},\
  \bibinfo {pages} {076402} (\bibinfo {year} {2007})}\BibitemShut {NoStop}%
\bibitem [{\citenamefont {Filippone}\ \emph {et~al.}(2018)\citenamefont
  {Filippone}, \citenamefont {Moca}, \citenamefont {Weichselbaum},
  \citenamefont {von Delft},\ and\ \citenamefont {Mora}}]{Filippone18}%
  \BibitemOpen
  \bibfield  {author} {\bibinfo {author} {\bibfnamefont {M.}~\bibnamefont
  {Filippone}}, \bibinfo {author} {\bibfnamefont {C.}~\bibnamefont {Moca}},
  \bibinfo {author} {\bibfnamefont {A.}~\bibnamefont {Weichselbaum}}, \bibinfo
  {author} {\bibfnamefont {J.}~\bibnamefont {von Delft}}, \ and\ \bibinfo
  {author} {\bibfnamefont {C.}~\bibnamefont {Mora}},\ }\bibfield  {title}
  {\enquote {\bibinfo {title} {At which magnetic field, exactly, does the kondo
  resonance begin to split? a {Fermi} liquid description of the low-energy
  properties of the {Anderson} model},}\ }\href {\doibase
  10.1103/PhysRevB.98.075404} {\bibfield  {journal} {\bibinfo  {journal} {Phys.
  Rev. B}\ }\textbf {\bibinfo {volume} {98}},\ \bibinfo {pages} {075404}
  (\bibinfo {year} {2018})}\BibitemShut {NoStop}%
\bibitem [{\citenamefont {Oguri}\ and\ \citenamefont
  {Hewson}(2018{\natexlab{a}})}]{Oguri17}%
  \BibitemOpen
  \bibfield  {author} {\bibinfo {author} {\bibfnamefont {A.}~\bibnamefont
  {Oguri}}\ and\ \bibinfo {author} {\bibfnamefont {A.~C.}\ \bibnamefont
  {Hewson}},\ }\bibfield  {title} {\enquote {\bibinfo {title} {Higher-order
  {Fermi}-liquid corrections for an {Anderson} impurity away from half
  filling},}\ }\href {\doibase 10.1103/PhysRevLett.120.126802} {\bibfield
  {journal} {\bibinfo  {journal} {Phys. Rev. Lett.}\ }\textbf {\bibinfo
  {volume} {120}},\ \bibinfo {pages} {126802} (\bibinfo {year}
  {2018}{\natexlab{a}})}\BibitemShut {NoStop}%
\bibitem [{\citenamefont {Oguri}\ and\ \citenamefont
  {Hewson}(2018{\natexlab{b}})}]{Oguri17_2}%
  \BibitemOpen
  \bibfield  {author} {\bibinfo {author} {\bibfnamefont {A.}~\bibnamefont
  {Oguri}}\ and\ \bibinfo {author} {\bibfnamefont {A.~C.}\ \bibnamefont
  {Hewson}},\ }\bibfield  {title} {\enquote {\bibinfo {title} {Higher-order
  {Fermi}-liquid corrections for an {Anderson} impurity away from half filling
  : Equilibrium properties},}\ }\href {\doibase 10.1103/PhysRevB.97.045406}
  {\bibfield  {journal} {\bibinfo  {journal} {Phys. Rev. B}\ }\textbf {\bibinfo
  {volume} {97}},\ \bibinfo {pages} {045406} (\bibinfo {year}
  {2018}{\natexlab{b}})}\BibitemShut {NoStop}%
\bibitem [{\citenamefont {Oguri}\ and\ \citenamefont
  {Hewson}(2018{\natexlab{c}})}]{Oguri17_3}%
  \BibitemOpen
  \bibfield  {author} {\bibinfo {author} {\bibfnamefont {A.}~\bibnamefont
  {Oguri}}\ and\ \bibinfo {author} {\bibfnamefont {A.~C.}\ \bibnamefont
  {Hewson}},\ }\bibfield  {title} {\enquote {\bibinfo {title} {Higher-order
  {Fermi}-liquid corrections for an {Anderson} impurity away from half filling:
  Nonequilibrium transport},}\ }\href {\doibase 10.1103/PhysRevB.97.035435}
  {\bibfield  {journal} {\bibinfo  {journal} {Phys. Rev. B}\ }\textbf {\bibinfo
  {volume} {97}},\ \bibinfo {pages} {035435} (\bibinfo {year}
  {2018}{\natexlab{c}})}\BibitemShut {NoStop}%
\bibitem [{\citenamefont {Iftikhar}\ \emph {et~al.}(2015)\citenamefont
  {Iftikhar}, \citenamefont {Jezouin}, \citenamefont {Anthore}, \citenamefont
  {Gennser}, \citenamefont {Parmentier}, \citenamefont {Cavanna},\ and\
  \citenamefont {Pierre}}]{Iftikhar15}%
  \BibitemOpen
  \bibfield  {author} {\bibinfo {author} {\bibfnamefont {Z.}~\bibnamefont
  {Iftikhar}}, \bibinfo {author} {\bibfnamefont {S.}~\bibnamefont {Jezouin}},
  \bibinfo {author} {\bibfnamefont {A.}~\bibnamefont {Anthore}}, \bibinfo
  {author} {\bibfnamefont {U.}~\bibnamefont {Gennser}}, \bibinfo {author}
  {\bibfnamefont {F.~D.}\ \bibnamefont {Parmentier}}, \bibinfo {author}
  {\bibfnamefont {A.}~\bibnamefont {Cavanna}}, \ and\ \bibinfo {author}
  {\bibfnamefont {F.}~\bibnamefont {Pierre}},\ }\bibfield  {title} {\enquote
  {\bibinfo {title} {Two-channel {Kondo} effect and renormalization flow with
  macroscopic quantum charge states},}\ }\href
  {http://dx.doi.org/10.1038/nature15384} {\bibfield  {journal} {\bibinfo
  {journal} {Nature}\ }\textbf {\bibinfo {volume} {526}},\ \bibinfo {pages}
  {233--236} (\bibinfo {year} {2015})}\BibitemShut {NoStop}%
\bibitem [{\citenamefont {Schwarz}\ \emph {et~al.}(2016)\citenamefont
  {Schwarz}, \citenamefont {Goldstein}, \citenamefont {Dorda}, \citenamefont
  {Arrigoni}, \citenamefont {Weichselbaum},\ and\ \citenamefont {von
  Delft}}]{Schwarz16}%
  \BibitemOpen
  \bibfield  {author} {\bibinfo {author} {\bibfnamefont {F.}~\bibnamefont
  {Schwarz}}, \bibinfo {author} {\bibfnamefont {M.}~\bibnamefont {Goldstein}},
  \bibinfo {author} {\bibfnamefont {A.}~\bibnamefont {Dorda}}, \bibinfo
  {author} {\bibfnamefont {E.}~\bibnamefont {Arrigoni}}, \bibinfo {author}
  {\bibfnamefont {A.}~\bibnamefont {Weichselbaum}}, \ and\ \bibinfo {author}
  {\bibfnamefont {J.}~\bibnamefont {von Delft}},\ }\bibfield  {title} {\enquote
  {\bibinfo {title} {Lindblad-driven discretized leads for nonequilibrium
  steady-state transport in quantum impurity models: Recovering the continuum
  limit},}\ }\href {\doibase 10.1103/PhysRevB.94.155142} {\bibfield  {journal}
  {\bibinfo  {journal} {Phys. Rev. B}\ }\textbf {\bibinfo {volume} {94}},\
  \bibinfo {pages} {155142} (\bibinfo {year} {2016})}\BibitemShut {NoStop}%
\bibitem [{\citenamefont {Werner}\ \emph {et~al.}(2016)\citenamefont {Werner},
  \citenamefont {Jaschke}, \citenamefont {Silvi}, \citenamefont {Kliesch},
  \citenamefont {Calarco}, \citenamefont {Eisert},\ and\ \citenamefont
  {Montangero}}]{Werner16}%
  \BibitemOpen
  \bibfield  {author} {\bibinfo {author} {\bibfnamefont {A.~H.}\ \bibnamefont
  {Werner}}, \bibinfo {author} {\bibfnamefont {D.}~\bibnamefont {Jaschke}},
  \bibinfo {author} {\bibfnamefont {P.}~\bibnamefont {Silvi}}, \bibinfo
  {author} {\bibfnamefont {M.}~\bibnamefont {Kliesch}}, \bibinfo {author}
  {\bibfnamefont {T.}~\bibnamefont {Calarco}}, \bibinfo {author} {\bibfnamefont
  {J.}~\bibnamefont {Eisert}}, \ and\ \bibinfo {author} {\bibfnamefont
  {S.}~\bibnamefont {Montangero}},\ }\bibfield  {title} {\enquote {\bibinfo
  {title} {Positive tensor network approach for simulating open quantum
  many-body systems},}\ }\href {\doibase 10.1103/PhysRevLett.116.237201}
  {\bibfield  {journal} {\bibinfo  {journal} {Phys. Rev. Lett.}\ }\textbf
  {\bibinfo {volume} {116}},\ \bibinfo {pages} {237201} (\bibinfo {year}
  {2016})}\BibitemShut {NoStop}%
\bibitem [{\citenamefont {Cui}\ \emph {et~al.}(2015)\citenamefont {Cui},
  \citenamefont {Cirac},\ and\ \citenamefont {Ba\~nuls}}]{Cui15}%
  \BibitemOpen
  \bibfield  {author} {\bibinfo {author} {\bibfnamefont {J.}~\bibnamefont
  {Cui}}, \bibinfo {author} {\bibfnamefont {J.~I.}\ \bibnamefont {Cirac}}, \
  and\ \bibinfo {author} {\bibfnamefont {M.~C.}\ \bibnamefont {Ba\~nuls}},\
  }\bibfield  {title} {\enquote {\bibinfo {title} {Variational matrix product
  operators for the steady state of dissipative quantum systems},}\ }\href
  {\doibase 10.1103/PhysRevLett.114.220601} {\bibfield  {journal} {\bibinfo
  {journal} {Phys. Rev. Lett.}\ }\textbf {\bibinfo {volume} {114}},\ \bibinfo
  {pages} {220601} (\bibinfo {year} {2015})}\BibitemShut {NoStop}%
\bibitem [{\citenamefont {Mascarenhas}\ \emph {et~al.}(2015)\citenamefont
  {Mascarenhas}, \citenamefont {Flayac},\ and\ \citenamefont
  {Savona}}]{Mascarenhas15}%
  \BibitemOpen
  \bibfield  {author} {\bibinfo {author} {\bibfnamefont {E.}~\bibnamefont
  {Mascarenhas}}, \bibinfo {author} {\bibfnamefont {H.}~\bibnamefont {Flayac}},
  \ and\ \bibinfo {author} {\bibfnamefont {V.}~\bibnamefont {Savona}},\
  }\bibfield  {title} {\enquote {\bibinfo {title} {Matrix-product-operator
  approach to the nonequilibrium steady state of driven-dissipative quantum
  arrays},}\ }\href {\doibase 10.1103/PhysRevA.92.022116} {\bibfield  {journal}
  {\bibinfo  {journal} {Phys. Rev. A}\ }\textbf {\bibinfo {volume} {92}},\
  \bibinfo {pages} {022116} (\bibinfo {year} {2015})}\BibitemShut {NoStop}%
\bibitem [{\citenamefont {Filippone}\ \emph {et~al.}(2017)\citenamefont
  {Filippone}, \citenamefont {Moca}, \citenamefont {Weichselbaum},
  \citenamefont {von Delft},\ and\ \citenamefont {Mora}}]{Filippone17arXiv}%
  \BibitemOpen
  \bibfield  {author} {\bibinfo {author} {\bibfnamefont {M.}~\bibnamefont
  {Filippone}}, \bibinfo {author} {\bibfnamefont {C.~P.}\ \bibnamefont {Moca}},
  \bibinfo {author} {\bibfnamefont {A.}~\bibnamefont {Weichselbaum}}, \bibinfo
  {author} {\bibfnamefont {J.}~\bibnamefont {von Delft}}, \ and\ \bibinfo
  {author} {\bibfnamefont {C.}~\bibnamefont {Mora}},\ }\bibfield  {title}
  {\enquote {\bibinfo {title} {At which magnetic field, exactly, does the
  {Kondo} resonance begin to split? a {Fermi} liquid description of the
  low-energy properties of the {Anderson} model},}\ }\href
  {https://arxiv.org/abs/1609.06165v2} {\bibfield  {journal} {\bibinfo
  {journal} {arXiv:1609.06165v3}\ } (\bibinfo {year} {2017})}\BibitemShut
  {NoStop}%
\end{thebibliography}%
 
\clearpage

\begin{center}
        \textbf{\large Supplementary material}
\end{center}

\setcounter{equation}{0}
\setcounter{figure}{0}
\renewcommand{\theequation}{S\arabic{equation}}
\renewcommand{\thefigure}{S\arabic{figure}}
\renewcommand{\thesection}{{S-\arabic{section}}} 

This supplementary material goes into the details of the numerical
calculations. In section \ref{sec: appendix Thermofield} we describe
the thermofield in more detail. In section \ref{sec: appendix
  discretization} we describe the discretization we use for the
leads. In section \ref{sec: details MPS} we give some technical
details for the MPS implementation.  Section \ref{sec: expectation
  values} discusses how to determine expectation values, and section
\ref{sec: accuracy} uses an example to illustrate the accuracy of our
approach. Section~\ref{sec: comparison} compares our
  results for the SIAM at high voltages to previous results, and
  section~\ref{sec-splitting-field} addresses the question of
  determining the magnetic field at which the Kondo resonance begins
  to split.

\section{The Thermofield Approach}\label{sec: appendix Thermofield}
\begin{figure}[b]
\includegraphics[width=\linewidth]{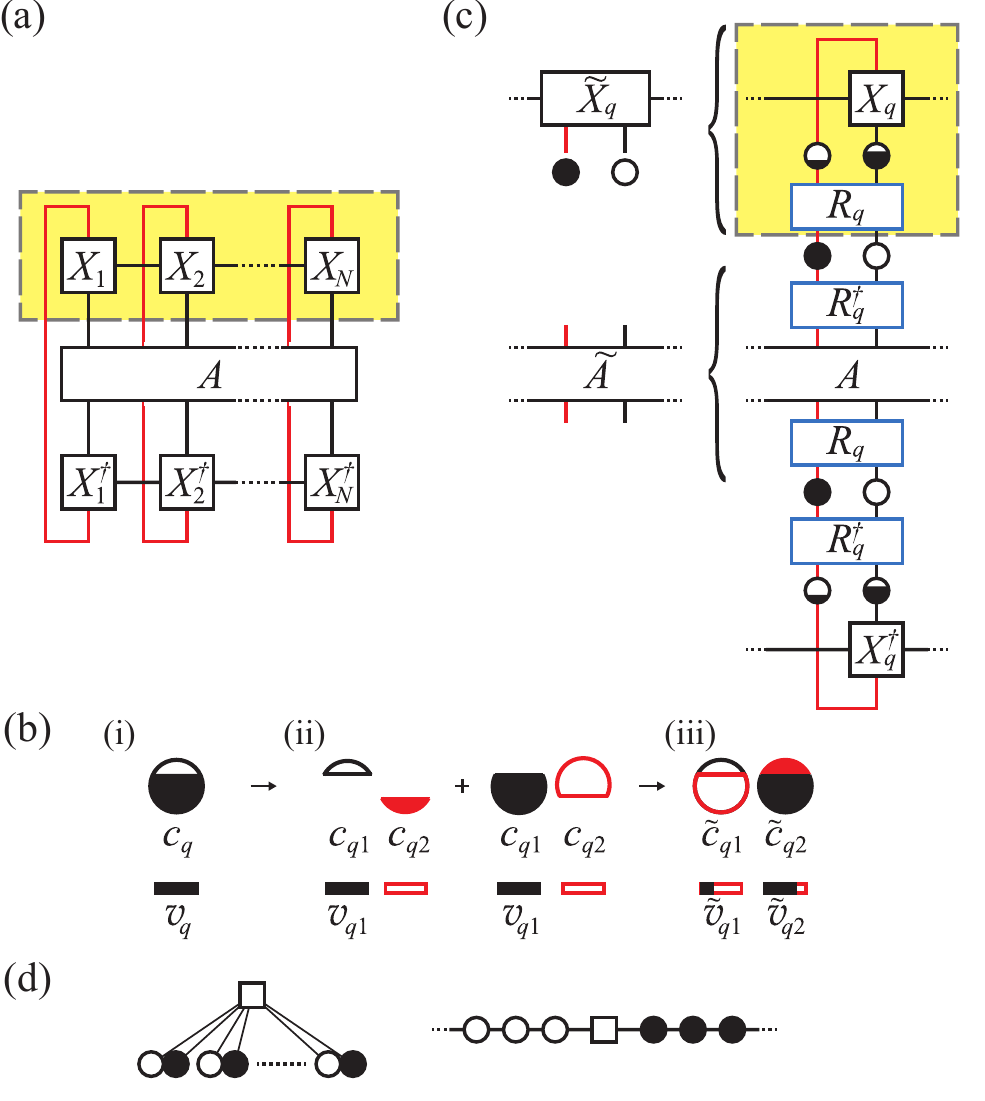}
\caption{
(a) Schematic MPS representation of
the expectation value $\braket{A}=\text{tr}{(\rho A)}$
rewritten in the form $\braket{\Omega|A|\Omega}$,
where the state $\ket{\Omega}$ with its physical
and auxiliary local modes is indicated by the dashed box.
(b) Starting from (i) a thermal level occupied with 
probability $f_q$ 
we represent the state (ii) as a
linear combination $\ket{\Omega}$ of states in which
the physical mode is empty or filled, weighting the
two contributions corresponding to the Fermi function.
We choose the auxiliary mode to be filled (empty)
when the physical mode is empty (filled)
[see Eq.\,(\ref{eq: app Omega})]. (iii) The rotation
$R_q$ in Eq.\,(\ref{eq: rot appendix}), combining
the physical mode ${c_{q1}}$ and the auxiliary mode
$c_{q2}$, yields modes that are empty or filled with 
probability one, but their coupling to the impurity
$\tilde{v}_{qi}$ depends on $f_q$.         
(c) Schematic depiction of the thermofield basis
transformation for a single fermionic level $q$.
Operators
$\tilde{A}$ 
act on 
the state $\ket{\Omega}$
represented in the new rotated basis 
consisting of ``holes'' and ``particles'' in terms of 
the tensors $\tilde{X}_q$.
(d) Both purification and local level rotation
are set up 
in the star geometry, where each ``free''
lead mode couples to the impurity only.
We then go over to the chain geometry by
tridiagonalizing the modes $\tilde{c}_{qi}$
such that the resulting Hamiltonian consists
of nearest-neighbor terms only.
We do this for the holes $\tilde{c}_{q1}$ and
the particles $\tilde{c}_{q2}$ separately.
Since both channels are product states
of either completely filled or completely
empty levels, a unitary one-particle basis
transformation, as provided by the tridiagonalization
performed separately within each channel only, necessarily preserves
this structure.
}
\label{fig: Purification}
\end{figure}

The thermofield approach  \cite{Takahashi75,Barnett87,Das00,deVega15}
 used in the main text 
is a convenient way
to represent a thermal state as a pure quantum state in an enlarged Hilbert space with the useful property that this pure state can be expressed as a simple product state.  Here, we summarize the analytic details of this approach. For a schematic depiction of its main steps, see Fig.\ \ref{fig: Purification}.

The density matrix of a thermal state is given by
\begin{flalign}
\rho =\tfrac{1}{Z(\beta)} 
e^{-\beta H} = \sum_n
\underbrace{\tfrac{e^{-\beta E_n}}{Z(\beta)}}_{\equiv \rho_n}
\ket{n}\bra{n}
\end{flalign}
with $\beta=1/T$, $Z(\beta)=\text{tr}\left(\e^{-\beta H}\right), \text{ and } \displaystyle{H\ket{n}=E_n\ket{n}}$.

Akin to purification \cite{Schollwoeck11},
one can represent this thermal state as pure
state $\ket{\Omega}$ in an enlarged Hilbert space:
one doubles the Hilbert space by introducing the auxiliary state space $\{\ket{n_2}\}$, which is a copy of the original Hilbert space $\{\ket{n}\}\equiv\{\ket{n_1}\}$ and defines, 
\begin{align}
\ket{\Omega} &=
\sum_{n_1,n_2}f_{n_1,n_2}(\beta)\ket{n_1}\otimes\ket{n_2}
\end{align}
such that the density matrix $\rho$ can be recovered as
\begin{align}
\label{eq: rho_1}
\notag\rho =& 
\mathrm{Tr}_\mathrm{aux} %
\left(\ket{\Omega}\bra{\Omega}\right) =
\sum_{n_2}\braket{n_2|\Omega}\braket{\Omega|n_2}
\\
=& \sum_{m_1, n_1} \underbrace{ \sum_{n_2}
  f^*_{m_1n_2}(\beta)f^{\vphantom{*}}_{n_1n_2}(\beta)
  }_{\equiv \rho_{n_1,m_1}} \ket{n_1}\bra{m_1}\,. &
\end{align}
Thermal equilibrium requires
\begin{align}
\label{eq:fn}
\rho_{n_1,m_1}
 =\tfrac{\e^{-\beta E_{n_1}}}{Z(\beta)}
  \delta^{\vphantom{*}}_{{m_1}{n_1}}
\,. 
\end{align}%
\Eq{eq: rho_1}
implies that the thermal expectation value of any operator $A$ is given by
\begin{align}
\braket{A}_\beta&=\braket{\Omega|A|\Omega}\,. &
\end{align}

For noninteracting systems we can look at each single fermionic mode $q$ separately with Hamiltonian \mbox{$H_q^\pdag=\varepsilon_q^\pdag c_q^\dagger c_q^\pdag$}. The orthonormal basis of our enlarged Hilbert space with modes $c_{q1}=c_q$ and $c_{q2}$ is given by:
\begin{align}
\label{eq: basis purification}\left\{\ket{0,0}_q, \ket{0,1}_q, \ket{1,0}_q, \ket{1,1}_q\right\}\,.
\end{align}
It follows from Eq.\,(\ref{eq:fn})
that the cumulative weight
of the first two states (where the physical mode is empty) is $(1-f_q)$ with $f_q=(1+\e^{\beta(\varepsilon_q-\mu_\alpha)})^{-1}$, while the weight of the other two (where the mode is filled) is $f_q$. 

Within the space of the four states in (\ref{eq: basis purification}) one can perform a rotation such that one of the new basis states carries the full weight in the thermal state, while the other three do not contribute. This can be exploited to represent $\ket{\Omega}$ as a simple product state. 
By choosing $\ds{f_{00}^{(q)}=f_{11}^{(q)}=0}$ (implying $\ds{f_{01}^{(q)}(\beta)=\sqrt{1-f_q}}$ and $\ds{f_{10}^{(q)}(\beta)=\sqrt{f_q}}$) and rotating such that $\ket{\Omega}=\prod_q\ket{\tilde{0},\tilde{1}}_q$,
we can ensure that this rotation preserves particle number conservation.

The rotated modes are of the form
\begin{align}
\label{eq: rot appendix}
\begin{pmatrix}\tilde{c}_{q1}\\\tilde{c}_{q2}\end{pmatrix}&=\begin{pmatrix*}[r]\cos\thetarot & -\sin\thetarot \\\sin\thetarot & \cos\thetarot\end{pmatrix*}\begin{pmatrix}{c}_{q1}\\{c}_{q2}\end{pmatrix}
\intertext{where the angle $\thetarot$ is defined by}
\begin{split}
\sin\thetarot&=f_{10}^{(q)}=\sqrt{f_q}\,,\\
\cos\thetarot&=f_{01}^{(q)}=\sqrt{1-f_q}\,. 
\end{split}
\end{align}
By construction, we then have 
\begin{align}
\label{eq: app Omega}
\ket{\Omega}&=\prod_q
\underbrace{
\left( \sqrt{1-f_q}\ket{0,1}_q+\sqrt{f_q}\ket{1,0}_q \right) 
}_{=: \, \ket{\tilde{0},\tilde{1}}_q}
\end{align}
and therefore
\begin{align}
\tilde{c}_{q1}^\pdag\ket{\Omega}&=\tilde{c}_{q2}^\dagger\ket{\Omega}=0\,.
\end{align}

Let us conclude with a few further remarks:
In the literature \cite{Das00,deVega15},
one typically transforms to a basis in which
$\ket{\Omega}$ is the vacuum of the enlarged
Hilbert space. This corresponds to the approach
presented here, but with the role of
$\tilde{c}_{q2}^\pdag$ and $\tilde{c}_{q2}^\dagger$
interchanged. In this case, the rotation in
Eq.\ (\ref{eq: rot appendix}) takes the
standard form of a Bogoliubov transformation.
Using this basis, it would not be necessary
to keep the rotated modes in separate channels
when going over to an MPS chain. However,
the mapping onto a single chain (i) does not
eliminate any degrees of freedom, and (ii) 
comes at the price of loosing particle
number conservation. Therefore, for the
sake of numerical efficiency, we preferred
to keep the two channels separate.
The only
drawback of the latter approach appears to be that particle
and hole excitations are locally separated
along the chain geometry which, eventually,
may make the accurate description of the
long-time behavior more challenging.

The thermofield approach is closely related to the purification
  approach often used in MPS studies of finite-temperature
  systems. 
In particular, both approaches 
involve doubling the degrees of freedom, 
introducing an auxiliary mode for each physical mode. But while the latter typically describes interacting systems, the thermofield approach corresponds to its application to noninteracting thermal leads.
In many applications of purification, the formulation is chosen such that auxiliary and physical modes are in the same state for the maximally entangled state at infinite temperature. For the thermal state of noninteracting leads at finite temperature, this would correspond to a choice of diagonal $f_{mn}^{(q)}$ in our statement below Eq.~(\ref{eq: basis purification}), such that
\begin{subequations}
\label{eq:purification}
\begin{align}
|\tilde{\Psi}\rangle =&
\sqrt{\rho_0} |0,0\rangle + \sqrt{\rho_1} |1,1\rangle
\,
\label{eq:purification:1}
\end{align}
for each single-particle lead level. In comparison to that, we exploit the freedom of unitary transformations in the auxiliary state space and use a number eigenstate instead,
\begin{align}
|\Psi\rangle =&
\sqrt{\rho_0} |0,1\rangle + \sqrt{\rho_1} |1,0\rangle
\, .
\label{eq:purification:2}
\end{align}
\end{subequations}
Evidently, Eq.~\eqref{eq:purification:2}
can be mapped onto Eq.~\eqref{eq:purification:1} 
by a particle-hole
transformation for the auxiliary degrees of freedom.
(In an MPS diagram such as 
\Fig{fig: Purification}(c), this would amount to flipping
 the direction of the arrow  of all lines \cite{Wb12_SUN} 
 representing auxiliary degrees of freedom.)
Since such a particle-hole transformation 
would map our $H_{\rm aux}$ onto $-H_{\rm aux}$,
the scheme used here is 
reminiscent of the purification scheme employed in 
\cite{Karrasch12}, who used opposite
signs for the physical and auxiliary mode
Hamiltonians in order to improve 
numerical efficiency.

Note also that in the present work we purify the thermal leads and do
not have an auxiliary degree of freedom for the impurity itself.  The
reason for this is simple: in the initial state we want to enforce a
specific thermal distribution on the occupation statistics of the
leads. This carries over to a specific connection between the
auxiliary and the physical degrees of freedom in the leads. In
contrast, the impurity can be in any state at the beginning of our
quench. In particular, one can choose the initial state of the
impurity such that the auxiliary mode for the impurity simply
decouples. Also the Hamiltonian dynamics does not connect the
auxiliary mode to the rest of the system, so we do not need to
describe the auxiliary degree of freedom for the impurity at any time.

Finally, we note that the present scheme
of simulating a thermal yet closed system can be
extended to open systems.
In a previous work \cite{Schwarz16}
we had
also introduced a lead representation in terms of
``holes'' and ``particles'', 
yet formulated a description of nonequilibrium
steady-state transport through a localized level
using Lindblad-driven discretized leads.
There we demonstrated, that such a Lindblad driving
in effect broadens the discrete levels of discretized
leads in such a way that they faithfully mimic
the properties of continuous leads. In the basis of
``holes'' and ``particles'' this Lindblad driving
takes a remarkably simple form and, in particular,
it is \textit{local} on the chain underlying the MPS.
By adding such a Lindblad driving to the time evolution,
it should be possible to describe even longer
time scales. However, the price one would have to pay,
is a
time evolution that is not described by
Hamiltonian dynamics but by a Lindblad equation.

\section{Log-linear Discretization}
\label{sec: appendix discretization}

We want to coarse-grain, i.e.\ discretize the full band of bandwidth
$[-D,D]$ into $N$ energy intervals $[E_n,E_{n+1}]$
in such a way that the width of the energy intervals
scales linearly within the transport window (TW)
$[-D^*,D^*]$ and logarithmically for energies outside,
with a sufficiently smooth transition between the
linear sector (lin-sector) and the logarithmic sector (log-sector).
Related ideas have been considered in \cite{Guttge2013,Schmitteckert10}.
The three relevant parameters for our discretization are:
(i) the level-spacing $\delta$ 
within the lin-sector;
(ii) the parameter $\Lambda >1$ defining the
logarithmic  discretization in the log-sector
(typically $\Lambda\gtrsim 2$; see below);
and (iii) the energy scale
$D^*$ at which the transition between the lin-sector and the
log-sector takes place. To construct such a log-linear
discretization we define a continuous function
$\mathcal{E}(x)$ which is evaluated at the points
$x_n=n+z$ with $n\in\mathbb{Z}$ and
$z\in[0,1)$ to obtain the energies
$E_n=\mathcal{E}(x_n)$. This function $\mathcal{E}(x)$
has to fulfill $\mathcal{E}(x+1)-\mathcal{E}(x)=\delta$
for $|\mathcal{E}(x+1)|<D^*$ and
$\frac{\mathcal{E}(x+1)}{\mathcal{E}(x)}=\Lambda$ %
$\bigl(\frac{\mathcal{E}(x)}{\mathcal{E}(x+1)}=\Lambda\bigr)$
for $\mathcal{E}(x)\gg D^*$ ($\mathcal{E}(x)\ll -D^*$),
respectively.
Furthermore, we demand the function and its first
derivative to be continuous.
We construct such a function by inserting a linear
section into the logarithmic discretization described
by the $\sinh()$ function,
\begin{align}
\mathcal{E}(x) =& \left\{\begin{array}{ll}
\delta\cdot x & \text{if } |x|\leq x^* \\
\delta\cdot\bigl(\tfrac{\sinh[(x\mp x^*)\log\Lambda]}{\log(\Lambda)}
\pm x^*\bigr) & \text{if }x\gtrless \pm x^*
\end{array}\right.
\end{align}
with $x^*=D^*/\delta$.
Fixing the three parameters $\delta$, $\Lambda$ and $D^*$
fully fixes the form of the function $\mathcal{E}(x)$.
The only free parameter left  is the parameter $z\in[0,1)$,
whose role is fully analogous to the $z$-shift
in NRG calculations \cite{Campo05,Zitko09}.
The outermost intervals are limited by the bandwidth $E_{1}=-D$, $E_{N+1}=D$. If one of these outermost intervals
gets narrow
compared to the adjoining interval, one can simply
join these two intervals into one for the sake
of energy scale separation within NRG. 

The discretization is therefore determined by
four parameters: $\Lambda$, $D^*$, $\delta$, and $z$.
The parameter $\Lambda$ characterizes the logarithmic
discretization for the log-sector. It has to be small enough
to capture the relevant high-energy physics, but
large enough to ensure energy scale separation
in the NRG calculation. For our calculations,
we typically choose $2\lesssim\Lambda\lesssim3$.
$D^*$ is the energy scale that defines the size
of the TW. If $T\lesssim V$, it is approximately
set by the chemical potential $V/2$.
If $T\gtrsim V$, temperature will define the size
of the TW and the edges of the window will be smeared out.
We chose $D^*$ as the energy at which the Fermi function
of the channel with positive chemical potential
$(\mu={V}/{2})$ has decreased to a value of $10^{-3}$,
implying $D^*={V}/{2}$ for $T=0$ and $D^*\approx 7T$
for $V\ll T$.
The level spacing $\delta$ in the lin-sector sets the
time-scale accessible by the quench calculations
before finite size effects get visible.
Typically, we set $\delta=D^*/20$, such that we have approximately forty energy intervals within the TW. In all our calculations, we used $z=0$.  

To each of the intervals $[E_n,E_{n+1}]$
we assign an energy $\varepsilon_n$ representing
the energy of the interval. In the context of NRG,
different methods have been developed to optimize
this energy \cite{Campo05,Zitko09}. Motivated by
Eq.\,(44) in Ref.\,[\onlinecite{Campo05}],
we choose a simplified version, namely
\begin{equation}
\label{eq: varepsilon:gen}
\varepsilon_n=\begin{cases}
\frac{E_{n+1}-E_{n}}{\ln\left({E_{n+1}}/{E_{n}}\right)},\quad&\text{if }\left|E_n\right|,\left|E_{n+1}\right|>D^* \\
\frac{1}{2}\left(E_n+E_{n+1}\right),\quad&\text{else.}
\end{cases}
\end{equation}
When $|E_n|$ approaches $|D^*|$ from above, our log-linear discretization approaches a linear discretization, with $E_{n+1}-E_n=\delta$. In this case,
\begin{align}
\label{eq: varepsilon}
\varepsilon_n &=
\tfrac{\delta}{\ln (1+\frac{\delta}{E_n} )}
\overset{\delta\ll E_n}{\approx}E_n+\tfrac{\delta}{2} 
\approx\tfrac{1}{2}\left(E_n+E_{n+1}\right),
\end{align}
which matches the definition of $\varepsilon_n$
for $|E_n|,|E_{n+1}|<D^*$
in \Eq{eq: varepsilon:gen}.
In this sense the smooth behavior of the energies $E_{n}$ defining the discretization intervals leads to a reasonably smooth transition from the log-sector to the lin-sector also in the energies $\varepsilon_n$. 

\section{Details on the MPS calculation}
\label{sec: details MPS}

All our MPS calculations were built on top of the
QSpace tensor library that can exploit abelian
as well as non-abelian symmetries on a generic
footing \cite{Wb11_rho,Wb12_FDM,Wb12_SUN}.
For the SIAM, standard particle-hole symmetry
is defined by the spinor $\hat{\psi^\dagger} \equiv
(c_\uparrow^\dagger, s c_\downarrow^\pdag)$, 
which  interchanges
holes and particles (up to a sign $s$) while simultaneously also
reverting spin 
$\sigma\in\{\up,\down\}$
\cite{Wb12_SUN}.
This symmetry acts independently of  
the SU(2) spin symmetry, and hence
is preserved even if $B\neq0$.
In our simulations, however, we 
only exploit U(1) spin and U(1) particle-hole
symmetry, since (i) we are also interested
in finite magnetic field $B$, which breaks
spin SU(2) symmetry, and (ii) finite bias voltage $V$
breaks particle-hole symmetry in the leads.

\subsection{The MPS geometry}
\begin{figure}
\centering
\includegraphics[width=\linewidth]{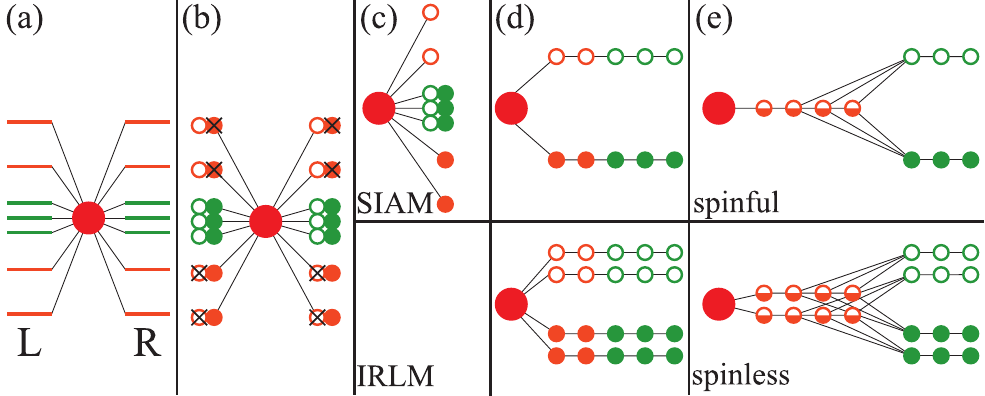}
\caption{
Sketch of the different discrete site geometries.
(a) We start with two channels $\alpha\in\{L,R\}$
in the star geometry, with the two colors representing
the log-sector and the lin-sector.
(b) Within the thermofield approach each level is
exactly represented 
by one ``hole'' and one ``particle''.
However, for the positive (negative) energies
in the log-sector the ``particles'' (``holes'') decouple
from the impurity due to $f_q=0$ ($1-f_q=0$),
respectively. 
(c) For the SIAM, only specific linear combinations
of left and right lead modes couple to the impurity,
while the corresponding orthogonal modes decouple see
Eqs.~\eqref{eq:Ctilde:ops}
(d) Tridiagonalizing ``holes'' and ``particles''
into separate channels, we get two channels
in the chain geometry for the SIAM (upper part) and
four in the IRLM (lower part), for which we still
distinguish between left and right leads.
The couplings in the log-sector for each channel
decay as $\Lambda^{-n}$ (e) Recombination of
holes and particles within log-sector into one channel
using another tridiagonalization
since for NRG it is unfavorable
to have ``holes'' and ``particles'' in separate
channels. The couplings
in this altered 
channel setup decay as $\Lambda^{-n/2}$, which resembles equilibrium NRG.
However, the first site of the lin-sector
in the chain geometry now couples to a range
of
sites of towards the end of the log-sector.
Nevertheless,
energy scale separation ensures that this
nonlocality is restricted to only a few
sites.
}
\label{fig: Channel Geometry}
\end{figure}

The starting point is the star geometry with the two leads,
$\alpha\in\{L,R\}$, discretized in energy with lead levels
$q=\{\alpha,(\sigma),k\}$, as depicted in
\Fig{fig: Channel Geometry}(a). Note that we do not
include the chemical potential into the energies $\varepsilon_q$.
Together with left/right symmetry for the leads,
this implies
$\varepsilon_{\alpha(\sigma) k}
=\varepsilon_{k}$.
In the thermofield approach the lead levels $q$ are doubled
and rotated to ``holes'' and ``particles'', represented
by the operators $\tilde{c}_{qi}$, as depicted in
\Fig{fig: Channel Geometry}(b).

\paragraph{Decoupling modes: }
For the positive (negative) high energies in the 
\mbox{log-sector} the ``particle'' modes $\tilde{c}_{q2}$
(the ``hole'' modes $\tilde{c}_{q1}$) 
are already
decoupled due to $f_q=0$ ($f_q=1$) without
any further rotation. Hence the doubling of levels
is not required there.

Furthermore, in the SIAM, we can combine the ``holes''
and ``particles'' separately
from the left lead with those from the right
lead into new modes,
\begin{subequations}
\label{eq:Ctilde:ops}
\begin{align}
\label{eq: combine channels SIAM}
\tilde{C}_{k\sigma i}&=\tfrac{1}{\mathcal{N}}
\sum_\alpha \tilde{v}_{\alpha k \sigma i}\tilde{c}_{\alpha k \sigma i}\ , \quad \bigl( \mathcal{N}^2 \equiv 
\sum_{\alpha'}\pdag |v_{\alpha' k\sigma i}|^2 \bigr)
\intertext{yielding the geometry in
\Fig{fig: Channel Geometry}(c). The modes orthogonal to these,}
\label{eq: combine channels SIAM 2}
\tilde{C}^{(\perp)}_{k\sigma i}
&= \tfrac{1}{\mathcal{N}}
   \left(
       \tilde{v}_{Lk\sigma i}^*\tilde{c}_{Rk\sigma i}
      -\tilde{v}_{Rk\sigma i}^*\tilde{c}_{Lk\sigma i}
   \right)\,,
\end{align}
\end{subequations}
decouple from the impurity. In matrix notation,
temporarily suppressing the global index set
$k\sigma i$ for readability, this can be written as
\begin{subequations}
\label{eq:Ctilde}
\begin{align}
\label{eq:Ctilde:1}
   \left(\begin{array}{l}
      \tilde{C}  \\
      \tilde{C}^{(\perp)} \\
   \end{array}\right)
   &= \tfrac{1}{\mathcal{N}} \left(\begin{array}{cc}
      \tilde{v}_{L} & \tilde{v}_{R} \\
     -\tilde{v}_{R}^\ast & \tilde{v}_{L}^\ast \\
   \end{array}\right)
   \left(\begin{array}{l}
      \tilde{c}_{L}  \\
      \tilde{c}_{R}  \\
   \end{array}\right)
\end{align}
with inverse relations,
\begin{align}
\label{eq:Ctilde:2}
   \left(\begin{array}{l}
      \tilde{c}_{L}  \\
      \tilde{c}_{R}  \\
   \end{array}\right)
   &= \tfrac{1}{\mathcal{N}} \left(\begin{array}{cc}
      \tilde{v}_{L}^\ast & -\tilde{v}_{R} \\
      \tilde{v}_{R}^\ast &  \tilde{v}_{L} \\
   \end{array}\right)
   \left(\begin{array}{l}
      \tilde{C}  \\
      \tilde{C}^{(\perp)} \\
   \end{array}\right)
\end{align}
\end{subequations}
The decoupling of the orthogonal modes
is in complete
analogy to standard equilibrium calculations
in the SIAM \cite{Bulla08}. 
In our setup it carries over to the nonequilibrium
situation, because the difference in the chemical
potential of the two physical leads is shifted
into the couplings $\tilde{v}_{qi}$. 
In the IRLM, this combination of left and right lead modes
is not possible, because the two leads couple to two
different impurity sites, in full analogy to standard
equilibrium calculations.

The above analysis leads to the remarkable
conclusion
that the numerical effort for the description
of the spinless IRLM is comparable to
that of the spinful SIAM.
The additional cost involved for the SIAM
for treating two states is compensated by the
simplification that left and right lead modes
can be combined because they couple to the
same impurity site.

\paragraph{Tridiagonalization: }
When going over to a chain geometry, the corresponding
tridiagonalization is performed for ``holes'' and
``particles'' independently (treating them as
different ``channels''), in order to maintain the
property that  the thermal state $\ket{\Omega}$
is a simple product state while also preserving
charge conservation:
if for the state $\ket{\Omega}$ a channel is
completely empty (filled) in the star geometry,
it will remain a completely empty (filled) channel
also in the chain geometry. 
For 
the IRLM, since the left and right leads
have to be represented as separate channels,
we tridiagonalize the modes $\tilde{c}_{qi}$ into the four channels $\{\alpha i\}$ with $\alpha\in\{L,R\}$ and $i\in\{1,2\}$ labeling ``holes'' and ``particles'', see lower part of Fig.\ \ref{fig: Channel Geometry}(d).  For the SIAM, in contrast, left and right leads are combined in the sense of equation (\ref{eq: combine channels SIAM}), so we separately tridiagonalize the ``holes'' ($\tilde{C}_{q,i=1}$) and the ``particles'' ($\tilde{C}_{q,i=2}$), see upper part of Fig.\ \ref{fig: Channel Geometry}(d). 

Due to energy scale separation, the first part of the
chain corresponds to the energy scales of the \mbox{log-sector},
while the later part of the chain represents the \mbox{lin-sector}. 
Instead of counting the exact number of sites
in the chain geometry, we identify the \mbox{log-sector}
by looking at the behavior of the hoppings
which decay exponentially in the log-sector and are all
of the same order in the lin-sector.
Due to the smoothened transition from the linear
to the logarithmic discretization also
the hopping matrix elements show a smooth crossover
from exponential decay to approaching a constant.
We define the log-sector on the chain as the part for which
(i) the  hoppings decay strongly enough
(the details of this condition slightly depend
on the number of many-particle states kept in
the NRG iterations) and (ii) the hoppings are
larger than the energy scale $D^*$ on which
transport takes place. By construction the two conditions
are roughly equivalent.
Note that for the ``holes'' (``particles'')
in the \mbox{log-sector} of the star geometry only
the positive (negative) energies contribute
to the hybridization. This translates into
a decay of the hoppings and on-site energies
scaling as $\Lambda^{-n}$ for the \mbox{log-sector}
on the chain.

\paragraph{Re-combining ``holes'' and ``particles''
in the log-sector:}

For the NRG calculation it is disadvantageous
to describe ``holes'' and ``particles''
in separate channels since particle-hole
excitations are sharply separated in terms of
the particle and the hole content along
the chain geometry. 
Consequently, we apply a further tridiagonalization
that remixes ``holes'' and ``particles''
of the \mbox{log-sector} into one channel, e.g.\ see
upper part of Fig.~\ref{fig: Channel Geometry}(e).
This then defines the renormalized impurity (RI).
For the IRLM, this subsequent
tridiagonalization is done for the left and right
lead separately, see lower part of
Fig.~\ref{fig: Channel Geometry}(e).
After this recombination the hoppings
in the channel(s) will decay as $\Lambda^{-n/2}$. 
The numerical complexity of the NRG calculation,
therefore, is comparable to that of a standard
equilibrium calculation in the sense that
we obtain the same number of numerical channels
(one spinful for the SIAM, two spinless for the IRLM)
and the same exponential decay in the energy scales.  
Note that the tridiagonalization combining ``holes''
and ``particles'' for the \mbox{log-sector} comes with
the caveat that it introduces a nonlocality in
the Hamiltonian: after this further tridiagonalization,
the first site in the \mbox{lin-sector} does not only couple
to the last site of the \mbox{log-sector} but 
rather to the last
few sites, see Fig.~\ref{fig: Channel Geometry}(e).
The corresponding hopping term is therefore subject
to truncation within the NRG iterations. However,
energy-scale separation ensures that this nonlocality
stretches only over a few sites, so the error
introduced by the truncation of this hopping
is considered minor. 

\paragraph{Remaining lin-sector:}

For the DMRG calculation we order the channels
such that the ``holes'' are on one side of the RI
and the ``particles'' on the other side,
see Fig.\ \ref{fig: Fig_DiscretizationAndMPS}. 
The local dimension of each chain element is given by $2^2=4$: in the SIAM this is due to the spin degree of freedom $\sigma$, in the IRLM it represents the remaining degree of freedom in the physical leads $\alpha$.

In case of the IRLM, where left and right lead
are kept separate, there is one further point worth noting:
at $T=0$, also in the \mbox{lin-sector} either the ``hole''
or the ``particle'' decouples from the impurity
for each lead level $q$. This implies that parts
of the remaining chains representing ``holes'' and
``particles'' in the \mbox{log-sector} of the chain
geometry decouple. This fact can be applied to
further reduce the numerical cost, even though
we have not done so here. 
It
stems from the fact that no purification procedure
is needed for $T=0$, and therefore
does not carry
over to $T>0$.

\subsection{Renormalized Impurity}
\label{sec:RI}

The log-sector traces out the
high-energy degrees of freedom at energies $E\gg T,V$.
Therefore the renormalized impurity represents
the low-energy many body basis that still spans
energies up to and beyond the transport window (TW)
set by $\max(T,V)$.
Typically we keep approximately $500$ to $700$ states
to describe this basis.
In the quench protocol, we can pick an arbitrary
pure state $\ket{\phi_\text{ini}}$ in this
effective low-energy space
as the initial state for the RI.
In order to avoid excess energy in the initial
state, we choose
the ground state of the log-sector.

If the ground state space is degenerate by symmetry,
picking a single individual state may artificially
break that symmetry. Therefore proper averaging
over degenerate state spaces is required,
either by actually running separate
simulations for each degenerate ground state,
or by simply exploiting the known effect of the
symmetry on the numerical result.
(This also applies to the case 
of quasi-degenerate ground states, e.g.\ when a
symmetry present in the Hamiltonian is
only weakly broken.)
Overall, note that degeneracy within the log-sector
is rather generic, 
since we choose to keep particle and hole
channels symmetric. Therefore
we combine the {\it same} 
number of ``hole'' and ``particle''
sites into the log-sector 
such that, including the impurity site, it always
contains an odd number of sites
[see \Fig{fig: Channel Geometry}].

For example, for the IRLM at particle-hole 
symmetry, the log-sector has a single zero-energy
level, $\varepsilon_c = 0$,
causing the ground state sector to be
two-fold degenerate. Since our NRG code exploits
abelian particle number conservation, we obtain two
ground states for the log-sector that are particle-number
eigenstates globally within the RI,
say $|G_1\rangle$ and $|G_2 \rangle$.
We can initialize our quench calculations by taking
$|\phi_{\rm ini}\rangle$ equal to either
$|G_1 \rangle$ or $|G_2 \rangle$.

Now, for a particle-hole-symmetric model involving
a zero-energy level coupled to an \textit{infinite}
bath, the local (e.g.\ thermal) occupancy is
$n_C = 1/2$. However, 
the initial 
local occupancies for the 
two number eigenstates above,
say $n_{C,i} = \langle G_i |\hat n_C |G_i\rangle$
(for $i=1,2$), are not necessarily equal.
In general, $n_{C,1}+n_{C,2}=1$, yet
$n_{C,1} \neq n_{C,2}$ due to finite-size effects (the log-sector involves only a \textit{finite}
number of bath levels). 
Correspondingly, during the post-quench time
evolution, only the average of the local occupancies,
$\langle n_C \rangle (t) \equiv \tfrac{1}{2}
(n_{C,1} + n_{C,2})(t) =1/2$,
throughout, whereas the local occupancies for the
two individual states, $n_{C,i} (t)$, reach the
value $1/2$ only in the asymptotic limit
$t \to \infty$ due to their hybridization
with the lin-sector.
In practice, by knowing the underlying symmetry
which enforces $n_{C,2}(t) = 1 - n_{C,1}(t)$,
the inialization of the quench may only include
e.g. $|G_1\rangle$, bearing in mind that the 
data must be symmetrized w.r.t.\ occupation.

Alternatively, one could construct linear
  combinations of $|G_{1,2}\rangle$, say $|G_\pm\rangle$, which are
  eigenstates of a particle-hole transformation with eigenvalues
  $\pm 1$, and which yield local occupancies,
  $n_{C,\pm} = \langle G_\pm | \hat n_C | G_\pm \rangle$, that by
  construction satisfy $n_{c,\pm} = 1/2$.  If we would initialize the
  quench by taking $|\phi_{\rm ini}\rangle$ equal to either
  $|G_+ \rangle$ or $|G_- \rangle$, then we would find $n_{C, \pm}(t) = 1/2$
  throughout the post-quench time evolution. However, since the
  post-quench time evolution conserves particle number within each
  particle-number eigensector, this strategy
  would be equivalent to averaging the result of two separate
  quenches, initialized with $|\phi_{\rm ini}\rangle$
equal to $|G_1 \rangle$ or $|G_2 \rangle$, respectively.

\subsection{Trotter time evolution}

\begin{figure}
\includegraphics[width=\linewidth]{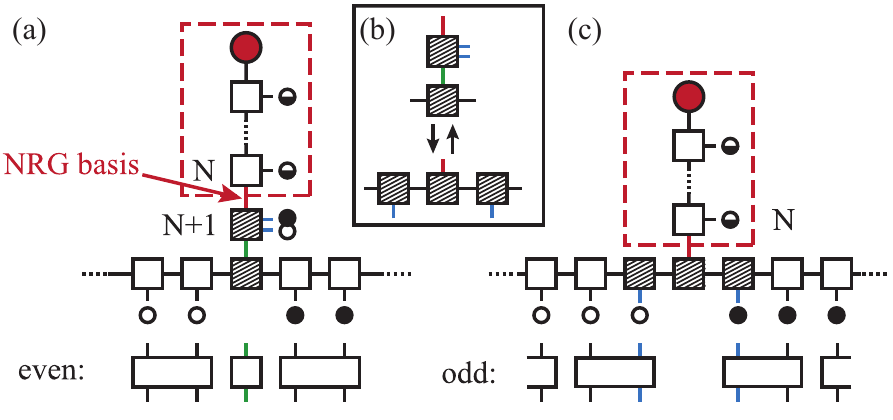}
\caption{
Sketch to illustrate
how NRG and DMRG are
combined in the Trotter time evolution.
(a) For the performance of the ``even'' time steps
we exponentiate the Hamiltonian of all NRG sites
plus one additional site in the sense of standard NRG.
For the ``odd'' time steps we rearrange the tensors
as depicted in (b) including fermionic swap gates
to bring the MPS into a form with local Trotter gates.
(c) The time evolution on the ``odd'' bonds
is then a standard tDMRG step.
The boxes at the bottom in both, (a) and (c),
indicate the Trotter gates to be applied.
}
\label{fig: Sketch MPS}
\end{figure}

The initial state is evolved in time,
$\ket{\Psi(t)}=\e^{-\ii Ht}\ket{\Psi(t=0)}$, using tDMRG \cite{Daley04,White04,Schollwoeck11}
with a standard second-order Trotter
decomposition for a time step $\tau$:
\begin{align}
\e^{-\ii H \tau } =& 
\e^{-\ii H_\text{o} \tau/2 }
\e^{-\ii H_\text{e} \tau }
\e^{-\ii H_\text{o} \tau/2 } +\mathcal{O}(\tau^3)\,,
\label{eq: Trotter decomposition}
\end{align}
where $H_\text{e}$ ($H_\text{o}$) includes all
``even'' (``odd'') bonds. The individual
terms in \Eq{eq: Trotter decomposition}
w.r.t.\ $H_\text{e}$ ($H_\text{o}$)
will be referred to as even (odd) Trotter
steps or even (odd) iterations, respectively.
The tensorial operations that are
performed in practice within the MPS setup,
are sketched  in \Fig{fig: Sketch MPS}.
The RI is described within a fixed effective
low-energy basis. The main idea is to use this
fixed basis as the local state space
of an 
MPS site in the center
when performing the Trotter time evolution.
However, when constructing the time evolution operator
that contains the coupling between the NRG sites and
the first of the remaining sites, one has to be careful
with the exponentiation of
the coupling term.
For this purpose, we need to consider
two subsequent NRG iterations, e.g.\ at Wilson
chain lengths $N$ and $N+1$, where site $N+1$
will be referred to as
flexible site.
These will be treated
differently in the even compared to the odd Trotter
steps 
(depending on the exact chain length, the notion of
``even'' and ``odd'' may need to be interchanged).
For the time steps which we call ``even'' in panel (a),
we exponentiate the full Hamiltonian of 
$N$ NRG sites plus the flexible site
($H^{\text{NRG}}_{N+1}$), 
yet excluding the coupling to the rest of the chain.
Therefore we {\it fully} associate the ``local''
Hamiltonian of the RI with even iterations which
is allowed within the Trotter setup.
Assuming that the Wilson chain length $N+1$ is
still within the realm of energy scale separation,
it can be dealt with in standard NRG manner.
In particular, it can be exactly diagonalized
in the expanded state space, including the
state space of the flexible site, followed
by simple exponentiation.
The couplings between the flexible site
and the subsequent sites, i.e.\ sites
$N+1$ and $N+2$, both left and right,
we reshape the tensors as depicted in
\Fig{fig: Sketch MPS}(b). 
Note that this requires 
fermionic swap gates \cite{Corboz10}
to account for the correct treatment
of fermionic signs.
After this reshaping the performance of the
``odd'' time steps is standard, as
sketched in \Fig{fig: Sketch MPS}(c). 

At time $t=0$, the RI is in its ground state,
while the leads are thermal. Since we are 
interested in the nonequilibrium steady-state
properties,
we do not switch on the coupling between RI
and thermal leads abruptly in our quench
protocol, as this would introduce undesirable
high-energy excitations into the system.
Instead, with adiabaticity in mind, we
turn on the coupling between RI
and thermal leads smoothly over a short
time interval.
The detailed form of this procedure should
not matter. In our calculation,
we ramp up the coupling $\eta$
between the RI and the thermal leads in a linear
fashion:
we use a time window of $t_\text{ramp}=2/D^*$
to $4/D^*$ and divide it into $N=10$ to $20$
equally spaced time intervals with
stepwise constant couplings,
$\eta(t_n)=\eta\frac{n}{N}$ where $n=1,\ldots,N$.

The size of the actual Trotter time step $\tau$ 
in equation (\ref{eq: Trotter decomposition})
should scale with $E_\text{trunc}^{-1}$,
with $E_\text{trunc}$ being the highest eigenenergy
of the truncated NRG basis
(or, if no NRG is required, the many-body
energy bandwidth, i.e.\ since all energy
scales are only moderately smaller as compared
to the bandwidth of the leads). In practice,
a prefactor of in the range $0.5$ to $1$ worked 
quite well.
In our calculation this energy $E_\text{trunc}$
typically is of the order $5D^*$ to $20D^*$.

When applying the Trotter gates, we keep
all singular values larger than some threshold
$\varepsilon_{(\mathrm{SVD})}$. 
Within our calculation this threshold varies
between $\varepsilon=2\cdot10^{-4}$ and $\varepsilon=10^{-3}$.
We time-evolve the system until a time
$t_{\rm max}$ at which a maximal bond dimension 
$D_\mathrm{max}$ is reached
in our MPS due to an increase in entanglement
entropy following the quench.
We used $D_\mathrm{max}$ up to $450$ in
our calculations.
The above parameters implied typical
accessible times in the post-ramp window
up to
$t_{\rm max}-t_{\rm ramp}>8/D^*$.
In case of $V\gg T$ this is equivalent to
$t_\text{max}>16/V$. Compared to an oscillation
period of $4\pi V^{-1}$ in the current (see below)
this range might seem rather small. 
However, typically these oscillations are
(a) strongly reduced in amplitude due to the
quasi-adiabatic quench protocol as described above,
and (b) in cases where the oscillations are
nevertheless still strong, i.e.\ at large voltages,
the accessible time window typically 
can be extended over many
periods.

\section{Expectation Values and Convergence}
\label{sec: expectation values}

\subsection{Current}

For the IRLM, the current through the central site
of the impurity can be defined by looking at the change
of the corresponding occupation number,
$\tfrac{d}{dt}\braket{n_C}$.
In the steady state this derivative should be zero,
of course, but we can identify the contribution,
$J_\alpha$, of the current flowing from lead
$\alpha$ into the dot from the formula
\begin{align}
0 = \tfrac{d}{dt}e\braket{{\hat{n}_C}}
  =  \sum_\alpha\underbrace{
  \tfrac{2\e}{\hbar} 
  \Im \bigl(t' \braket{d_C^\dagger d^\pdag_\alpha} \bigr)
  }_{\equiv J_\alpha}.
\end{align}

In the SIAM we combine the modes of the
left and right channels as given in
Eqs.\ (\ref{eq:Ctilde:ops}).
Still, it is possible to deduce the current
from the change of occupation
$\frac{d}{dt}\braket{\hat{n}_{d\sigma}}$
at the central site:
\begin{align}
\notag J_{\alpha\sigma} &= \tfrac{2\e}{\hbar}
\sum_{k}\Im
  \bigl(v_{q}\braket{d^\dagger_\sigma c_{q}^\pdag} \bigr) 
= \tfrac{2\e}{\hbar} \sum_{k i}\Im
  \bigl( \tilde{v}_{qi}
    \braket{d^\dagger_\sigma \tilde{c}_{qi}^\pdag}
  \bigr) \\
\label{eq: current SIAM}
&= \tfrac{2e}{\hbar} \sum_{k i}
   \tfrac{|\tilde{v}_{qi}|^2}{
      \sqrt{\sum_{\alpha'} \tilde{v}^2_{\alpha' k\sigma i}}
   }
   \Im\bigl(
     \braket{d^\dagger_\sigma \tilde{C}^\pdag_{k\sigma i}}
   \bigr)\,. 
\end{align}
where we used \Eq{eq:Ctilde:2}, 
$\tilde{c}_{\alpha k \sigma i}
= \tfrac{\tilde{v}_{\alpha k \sigma i}^\ast}{\mathcal{N}} 
\tilde{C}_{k\sigma i} + ...
\tilde{C}^{(\perp)}_{k\sigma i}$,
together with the fact that
the mode $\tilde{C}^{(\perp)}_{k\sigma i}$ 
decouples from the impurity and therefore
$\braket{d^\dagger_\sigma \tilde{C}^{(\perp)}_{k\sigma i}}=0$.
The chain operators underlying the MPS $f_{n\sigma (i)}$
are related to the modes $\tilde{C}_{k\sigma i}$ by a unitary
transformation, which includes the  mapping of ``holes'' and ``particles''
onto a chain and the re-combination of channels within the RI.
The expectation values
$\braket{d^\dagger_\sigma \tilde{C}^\pdag_{k\sigma i}}$
can therefore be determined by calculating the expectation
values $\braket{d^\dagger f_{n\sigma (i)}}$ for all chain
sites $n$. 
For the SIAM, the current can further be divided
into different spin contributions $J_{\alpha\sigma}$.

Interestingly, in most cases the
\textit{symmetrized} current
\begin{align}
J_{(\sigma)}&=\tfrac{1}{2}\left(J_{L(\sigma)}-J_{R(\sigma)}\right)\,.
\label{eq:Javg}
\end{align}
converges much faster than $J_{L(\sigma)}$ and $J_{R(\sigma)}$ separately
[see discussion of \Fig{fig: convergence}(h)
below for details]. For the SIAM, a similar statement holds when averaging over spin instead of averaging over channels. In practice, we take the mean over both by defining
\begin{align}
J&=\left(J_{\up}+J_{\down}\right)=\tfrac{1}{2}\left(J_{L\up}-J_{R\up}+J_{L\down}-J_{R\down}\right)
\end{align}

We define the value of the steady-state current $J(V)$ by taking the mean over the last part of $J_V(t)$, where the current is converged to its steady-state value. If the oscillations are pronounced, we take the mean over a time window, which equals an integer number of periods, in many cases simply the last period.
The conductance is obtained from
\begin{align}
g(V^*)=\frac{J(V_1)-J(V_2)}{V_1-V_2}\left(\frac{2e^2}{h}\right)^{-1}
\end{align}
with $V^*=\frac{1}{2}\left(V_1+V_2\right)$, and $V_1$ and $V_2$ close to each other, where we average $J_{V_1}(t)$ and $J_{V_2}(t)$ over similar time windows.

\subsection{Dot Occupation}

The occupation of the impurity in the SIAM,
as well as the occupation of the central
site of the impurity for the IRLM are
of physical relevance. Their 
time evolution is
related to that of the current via 
\begin{align}
\label{eq: connection N_d J_alpha}
 \tfrac{d}{dt}e\braket{{n}_{C/d}(t)} = J_L(t)+J_R(t)
\end{align}
In the present work, we focus on the particle-hole
symmetric point. Because of this symmetry we expect
the steady-state value of $n_{C/{d}}$ to be
independent of voltage and given by $n_C=\tfrac{1}{2}$
in the IRLM and $n_d=n_{d\up}+n_{d\down}=1$ in the
SIAM.  
The magnetization $M=\frac{1}{2}(n_{d\up}-n_{d\down})$,
however, is a nontrivial function of voltage
and magnetic field. 

\subsection{Long-time convergence after the quench}\label{sec: convergence}

\begin{figure*}[t]
\centering
\includegraphics[width=1\linewidth]{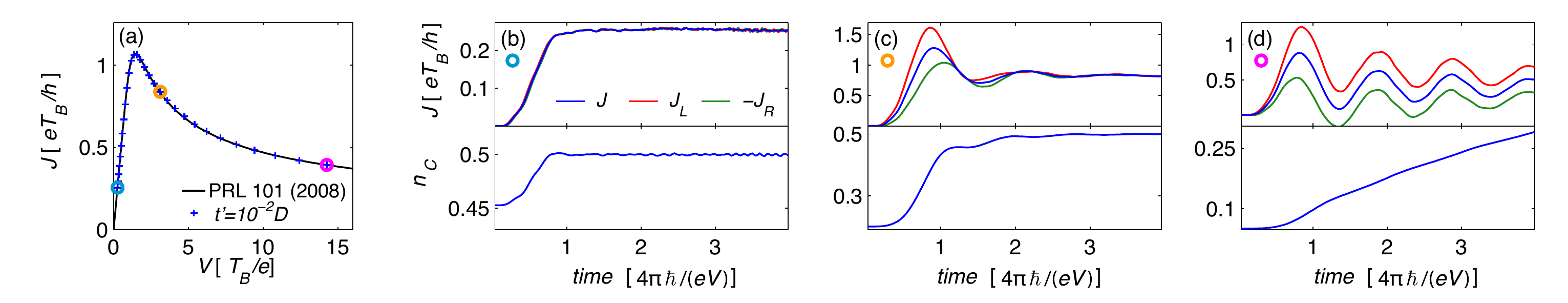}
\includegraphics[width=1\linewidth]{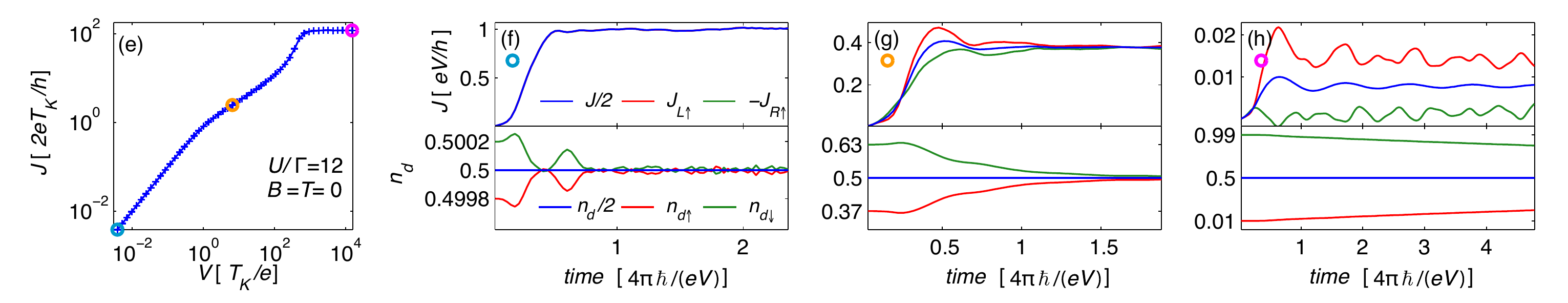}
\caption{
Upper panels: convergence in the IRLM ---
Panel (a) replots the data set for $t'=10^{-2}D$
in \Fig{fig: IRLM Scaling} of the main text.
Panels (b-d) show the time dependence of the currents $J_L$, $-J_R$ and $J =(J_L - J_R)/2$ and the dot occupation $n_C$, for the three different voltage values marked by circles in panel (a), respectively. For the lowest voltage which is still in the linear response regime, we find nice convergence in the dot occupation and the current (panel b). With increasing voltage, all three currents develop increasingly strong oscillations,  with a period of $4\pi/V$, as expected (panels c,d). For the largest voltage we do not find convergence in $n_C$ (panel d). This reflects in the fact that also $J_L$ and $J_R$ are not yet converged. However, the symmetrized current $J$ (blue line) does oscillate around a well-defined mean value.
Lower panels: convergence in the SIAM --- Panel (e) replots the data for $U=12\Gamma$ and $T=0$ in \Fig{fig: GoverTV} (a) of the
main text. Panels (f-h) show the behavior of $J_{L\up}$ and $J_{R\up}$, $J$, and $n_{d\sigma}$ (at  $T=B=0$) for the voltage values marked in circles in panel (e), respectively.  The current for the down-spin is not shown, since $J_{L\down}\approx -J_{R\up}$ and $J_{R\down}\approx -J_{L\up}$. The total dot occupation $n_d$ is equal to 1 in the beginning and remains so throughout.  However, for large voltages the numerically accessible time window is too short to find convergence for the spin-resolved occupations $n_{d \uparrow}$ and $n_{d \downarrow}$. In panel (h), the left and right components of the current (red and green lines) show seemingly irregular oscillations; these arising from a combination of large voltage and the finite level spacing in the lin-sector. The level-spacing effect cancels out, however, for the symmetrized current, $J= (J_L - J_R)/2$ (blue line), which shows regular oscillations with the expected period of $4 \pi/V$.}   
\label{fig: convergence}
\end{figure*}

By definition,
in the nonequilibriuim steady state (NESS) all
expectation values are converged in the sense that they do not change with time. However, we are limited to a finite time window and cannot fully reach this point. In this section, we discuss this aspect in more detail based on the behavior of the symmetrized current $J$, the currents from the left and right leads $J_{\alpha(\sigma)}$, and the (spin-resolved) dot occupation $n_{C}$ or $n_{d(\sigma)}$. 

As
explained above, 
our initial state breaks certain symmetries. However, as we assume the steady state to be unambiguous, we expect it to obey the symmetries of the Hamiltonian.

For the IRLM we have done our calculations at the
particle-hole symmetric point. We therefore expect
$n_C=1/2$ in the steady state. And, if the dot
occupation is converged, one finds $J_L=-J_R$ 
because of Eq.\ (\ref{eq: connection N_d J_alpha}).
This is, indeed, what we find for low voltages,
see \Fig{fig: convergence}(b). For higher voltages,
however, we do not see full convergence in $n_C$,
see \Fig{fig: convergence}(d).
Consequently, also the 
currents are not converged,
so we do not find $J_L=-J_R$.
However, the \textit{symmetrized} current $J$
is converged, except for oscillations around a
well-defined mean value. These oscillations do have
the expected period of $\frac{4\pi}{V}$
\cite{Schneider06}, 
and the amplitudes decay rapidly.
The initial state breaks particle-hole symmetry
as explained above. This symmetry breaking is
more pronounced for 
shorter 
NRG Wilson chains.
This is the reason why for small voltages
(for which the TW is small so that the NRG Wilson
chain is long) we already start with
$n_{d\sigma}(t=0)\approx\frac{1}{2}$ while for
high voltages (for which the TW is large and
the NRG Wilson chain is short) the symmetry
breaking in the beginning is very strong.

\begin{figure*}[t]
\centering
\includegraphics[width=1\linewidth]{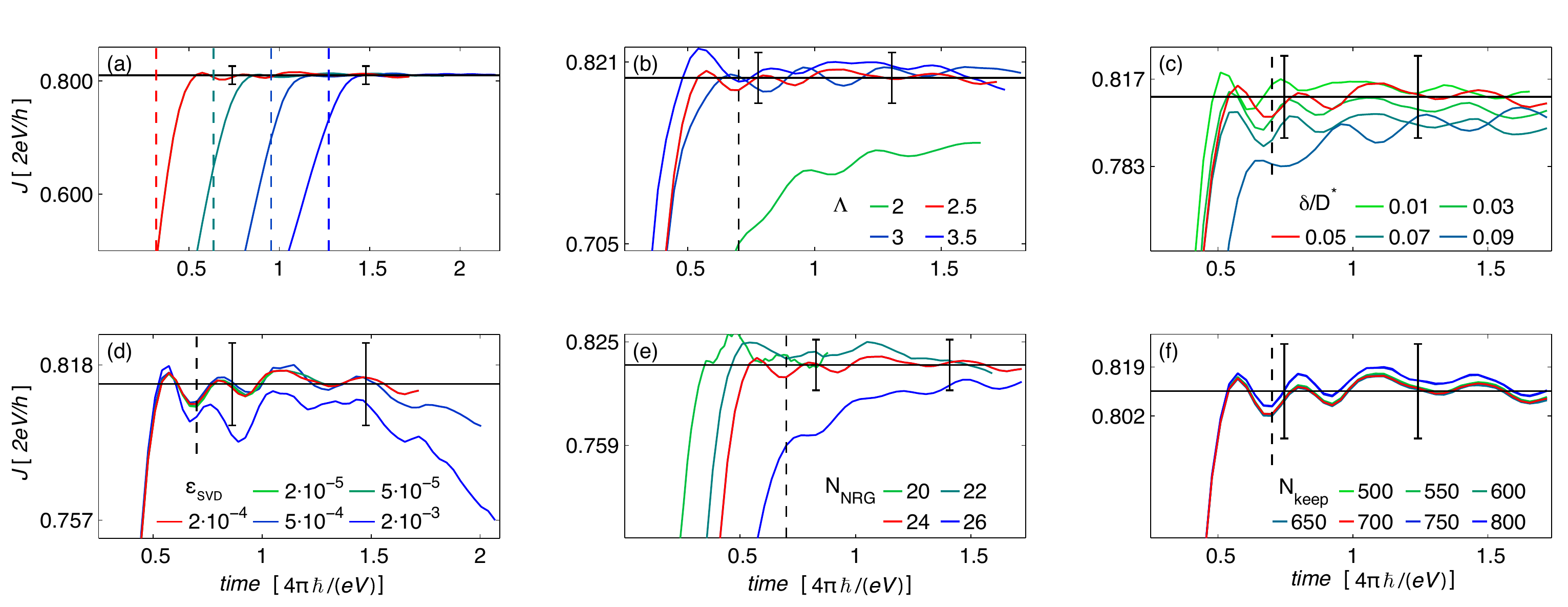}
\caption{
Illustration of the numerical accuracy
using the example of $V = T_K$ with the parameters
as in \Fig{fig: GoverTV}(a), with $U/\Gamma=12$.
In each of the panels the red curve corresponds
to the parameters typically used for our calculations
and the ``error bars'' indicate a relative range of
$\pm2\%$ around the mean.
Panel (a) shows $J(t)$ where the coupling between RI
and leads is turned on quasi-adiabatically over
time windows of four different widths.
All curves approach the same steady-state value.
In (b) and (c) the discretization
parameters $\Lambda$ and $\delta$ are varied.
In (d) different thresholds,
$\varepsilon_\mathrm{SVD}$,
are used for the SVD truncation in the tDMRG quench.
In (e) the number of sites treated with NRG is
changed (and therefore the number of sites treated with tDMRG
is changed accordingly). And finally in (f)
we use different numbers of kept states in the
effective NRG basis for the renormalized impurity.
}
\label{fig: errorbars}
\end{figure*}

Analogous considerations apply for the SIAM.
We numerically observe 
 the behaviour
\begin{align}
\label{eq:symmetries-for-current}
J_{L\sigma}(z) \approx -J_{R,-\sigma}(t)
\end{align}
and 
$n_d(t)=n_{d\up}(t)+n_{d\down}(t) \approx 1$
for all times $t$,
reflecting particle-hole and left-right symmetry
(here $-\sigma$ stands for reverted spin $\sigma$).
However, by choosing a specific initial pre-quench state out of a degenerate ground state multiplet, this breaks the spin symmetry, and hence we find
$n_{d \uparrow} (t) \neq n_{d \downarrow}(t)$, even for $B = 0$.
The effect of this symmetry breaking is largest for high voltages.
Whereas for small voltages we do find convergence in the dot
occupation [e.g. see \Fig{fig: convergence}(f)], for high voltages our
numerically accessible time window is too small to see convergence
[\Fig{fig: convergence}(h)].  Moreover, for large voltages the
spin-resolved currents $J_{L\sigma}$ and $J_{R \sigma}$ show seemingly
irregular oscillations, as seen in \Fig{fig: convergence}(h).
 A Fourier-transform analysis (not shown) reveals that the
  oscillations in $J_{\alpha \sigma}(t)$ 
  have several characteristic frequencies, one being
  $\frac{V}{4 \pi}$ (as expected from \cite{Wang10}), the others being
  the energies representing the intervals in the
  log-sector closest to $D^\ast$, which was chosen
   $D^\ast = V/2$ here.
  Thus, at large voltages the post-quench dynamics become
  sensitive to the rather crude discretization in the log-sector,
  causing the seemingly irregular oscillations in the spin-resolved
  currents at large voltages.  This suggests that the strength of
  these discretization-related oscillations could be reduced, if
  desired, by using a slower ramp for the quench (i.e.\ a larger
  ramping time $t_{\rm ramp}$), or by reducing the size of the
  log-sector (i.e.\ increasing $D^\ast$, while keeping the level
  spacing $\delta$ for the lin-sector fixed). In practice, though, we
  found this to be unnecessary, since the discretization-related
  oscillations cancel in the left-right symmetrized current: 
  $J = (J_L - J_R)/2$ shows only regular oscillations around a
  well-defined mean value [\Fig{fig: convergence}(h)] 
with the expected time-period of
  $\frac{4\pi}{V}$ \cite{Wang10}, similar to those found for the
  IRLM. We suspect that this cancellation of discretization-related 
  oscillations occurs because our treatment
  of the leads respects left-right symmetry, both regarding 
their discretization [see Fig.~\ref{fig: Channel Geometry}(d,e)] 
and when turning on the coupling
  between the log- and lin-sectors during the quench.

In the case of finite magnetic field in the SIAM,
we do not have spin symmetry. In particular,
we expect $n_{d\up}\neq n_{d\down}$,
even in the steady state. The exact NESS values
of $n_{d\sigma}$ are nontrivial and depend
on voltage. However for large values of $V$,
we are not able to see convergence in these
occupations, analogously to \Fig{fig: convergence}(h).
Still, it is in principle
possible to predict the NESS
occupation by extrapolating the 
data available within the accessible time window,
e.g.\ using linear prediction \cite{Barthel09}.

\section{Numerical accuracy}
\label{sec: accuracy}

Our approach treats the many-particle aspect of 
impurity models nonperturpatively. However, of course,
the numerics contains approximations such as the
discretization of the lead into a finite number
of energy intervals, the truncation of states
within the NRG, and the truncation of the MPS
within the tDMRG time evolution. A further error
arises from the fact that we have to take the mean
over a curve $J(t)$ that often still oscillates
over a well-converged mean value.
Therefore it is difficult to give a precise value
for our error. However, we can provide an 
estimate for the error bar. For the case of
the current, it is approximately $\pm3$\%, throughout,
which at times may be considered conservative. 

To illustrate this statement we go into
more detail for the curve $J(t)$ for the parameters
used in Fig.~\ref{fig: GoverTV}(a) with $U/\Gamma=12$
at $V\approx T_K$: 
Fig.~\ref{fig: errorbars} shows the behavior of $J(t)$
when varying various different numerical parameters,
such as discretization and truncation parameters.
In each of the panels the red curve was obtained
from the parameter choices typically used
in our numerics. This curve is identical
in each of the panels. 
The black horizontal line
shows the mean value obtained for times after
the vertical dashed black marker.
The ``error bars'', for convenience, indicate a range
of $\pm 2$\% 
around the mean value.
The essential 
message from all these plots is
that even though our results do show slight dependence
on the various numerical parameters that were
varied here, this dependence is small, and within
the stated error bars of $\lesssim 2$ to $3\%$.
Depending on the precise
parameters the curves in some cases wiggle more
strongly, or for higher voltages show stronger
oscillations. In this cases, the error
is closer to the upper end of the estimated error range. 
Looking at the comparison of $U=0$ with exact results and the comparison of $g(T,0)$ with NRG values in Fig.~\ref{fig: GoverTV}(a), confirms this estimate for our error bar.

\begin{figure}[tb!]
\includegraphics[width=\linewidth]{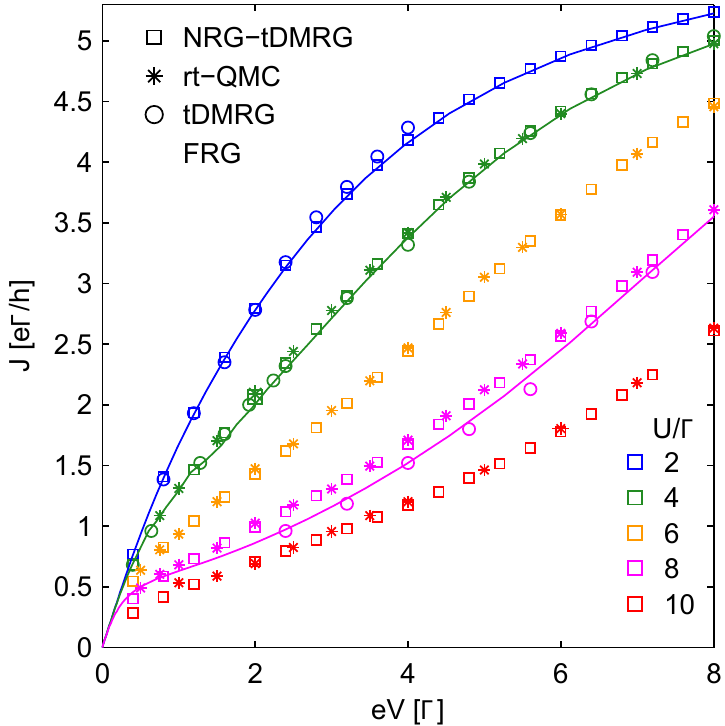}
\caption{Current in the SIAM as a function of voltage in the
    high-voltage regime, $V\gtrsim \Gamma$, for different values
    of $U/\Gamma$: we compare results obtained with our method
    (NRG-tDMRG) to results from rt-QMC, previous tDMRG calculations,
    and FRG (see Refs.\,[\onlinecite{Eckel10,Werner10}] for
    details). For a range of $U/\Gamma$ values our results nicely
    agree with previous results. }
\label{fig: comparison}
\end{figure}

\section{Comparison to other methods}
\label{sec: comparison}
In Ref.\,[\onlinecite{Eckel10}] previous tDMRG quench results on the
high-voltage regime of the SIAM are compared to results obtained via
the functional renormalization group (FRG) and real-time quantum Monte
Carlo (rt-QMC), see Refs.\,[\onlinecite{Eckel10,Werner10}] for details
on the different methods. Fig.\,\ref{fig: comparison} shows the data
of Fig.\,2 in Ref.\,[\onlinecite{Eckel10}] together with further
rt-QMC results taken from Ref.\,[\onlinecite{Werner10}]. For
comparison, we here also include results obtained in our NRG-tDMRG
quench setup. For all parameters our data nicely agree with the rt-QMC
data. For $U/\Gamma=8$, tDMRG and FRG slightly differ from the rt-QMC
results (and thus also from our results). This has already been
discussed in Ref.\,[\onlinecite{Eckel10}]. Note, however, that the
parameter regimes of these reference systems stayed far away from
low-energy Kondo scales since for the larger values of $U/\Gamma$ the
described regime corresponds to $V\gg T_K$, while the small values of
$U/\Gamma$ do not describe the Kondo limit.  

\begin{figure}[tb!]
\includegraphics[width=\linewidth]{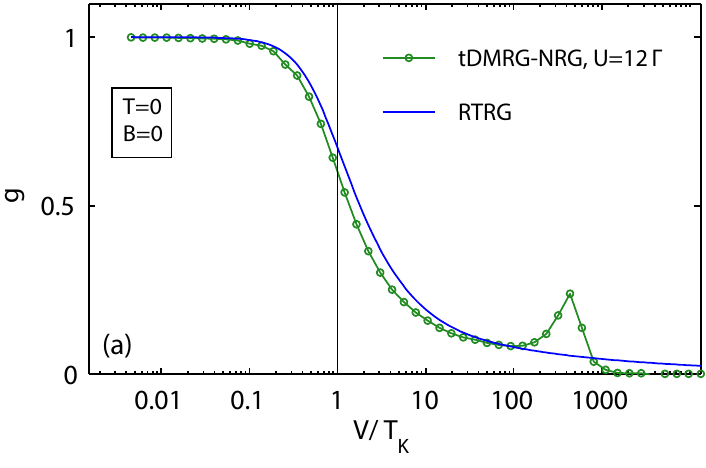}
\includegraphics[width=\linewidth]{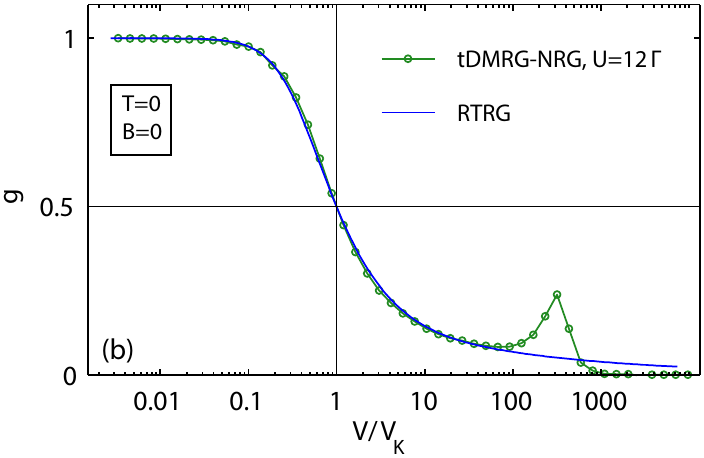}
\caption{ (a) Comparison of $g^{T=0}(V)$ vs.\ $V/T_K$ on 
a logarithmic scale, computed at $B=0$ in the Kondo limit. 
The data points (circles) show the NRG-tDMRG result for $U/\Gamma=12$,
    replotting the corresponding curve from Fig.\,3(a) of the main
    text. The solid curve shows the RTRG results of 
Pletyukhov and Schoeller [\onlinecite{Pletyukhov12}] for the Kondo model, 
plotted using  Eqs.\,(\ref{eq: RTRG
      result}). The small high-energy peak of the 
tDMRG-NRG curve reflects charge fluctuations not 
captured by the Kondo model. (b) Same data, but now
plotted vs.\ $V/V_K$.
}
\vspace{-\baselineskip}
\label{fig: S6b compare RTRG}
\end{figure}

We also compare our results  for the nonlinear conductance to 
those obtained by Pletyukhov and Schoeller   for the Kondo model using 
the real-time renormalization group (RTRG) in
Ref.\,[\onlinecite{Pletyukhov12}]. 
They found that the temperature and voltage scales
at which the conductance reaches $\frac{1}{2}$, defined
via 
\begin{align}
g^{V=0} (T=T_K)=\tfrac{1}{2}, \qquad 
g^{T=0}(V=V_K)=\tfrac{1}{2},
\end{align}
differ, 
with  
$V_K/T_K \approx 1.8$. 
(They use
the notation $T_K^\ast = T_K$ and $T_K^{\ast\ast}= V_K$.)
Their result for the nonlinear conductance 
can be fit well using the trial function
\begin{align} 
\nonumber
g^{T=0}_{\rm RTRG}(V)&
\approx \left\{1+\left[V/T_K'(x)\right]^2\right\}^{-s}, 
\quad x=V/V_K , \\
T_K'(x)&=T_K^{\ast\ast}\left(\frac{1-b+b x^{s'}}{2^{\frac{1}{s}}-1}\right)^{\frac{1}{2}} \, , \label{eq: RTRG result}
\end{align}
using $s=0.32$, $b= 0.05$ and $s'=1.26$.
Assuming that our data for $U/\Gamma=12$ in Fig.~3 of the main text is
deep in the Kondo limit, we compare our data for $g^{T=0}(V)$ vs.\
$V/T_K$ to theirs in Fig.~\ref{fig: S6b compare RTRG}(a).
Our curve for the nonlinear conductance has a
shape similar to theirs, but differs quantitatively in that it bends
downward somewhat more quickly.   Another way to quantify the difference
is to compare the predictions for the conductance at the voltage
$V=T_K$.  As mentioned in the main text, our calculations yield
$g(V=T_K)\approx0.6$, whereas RTRG predicts a value of approximately
$2/3$.

Despite this discrepency, we note that if both our and the RTRG conductance
curves are plotted versus 
$V/V_K$, 
thus making the comparison independent
of the finite-temperature, equilibrium scale $T_K$, 
the two curves almost coincide over
a wide range of
$V/V_K$ values, see Fig.~\ref{fig: S6b compare RTRG}(b). 
This suggests that the reason for the discrepancy
in Fig.~\ref{fig: S6b compare RTRG}(a) is that
the RTRG approach has an inaccuracy of a few percent 
in its determination of the ratio 
$V_K/T_K$.

\section{Splitting field in the SIAM}
\label{sec-splitting-field}
\begin{figure}
\includegraphics[width=\linewidth]{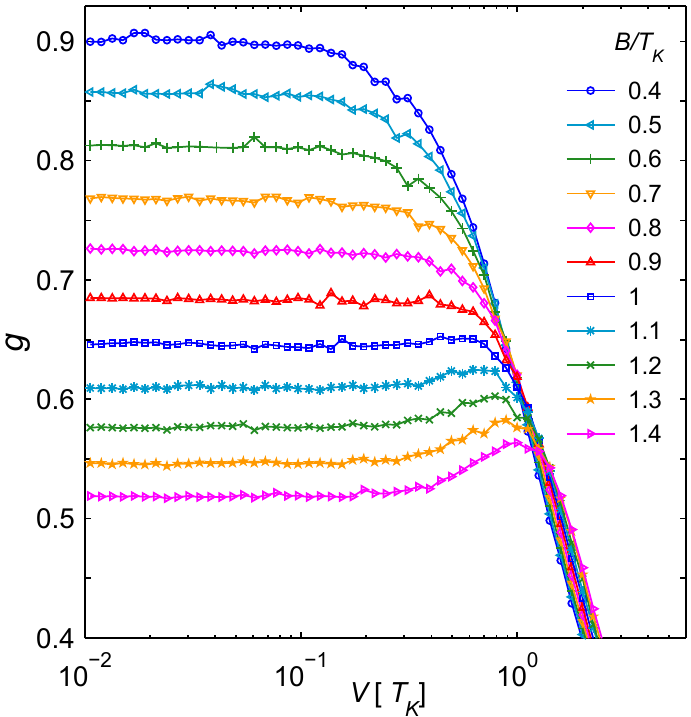}
\caption{Conductance as a function of voltage at $T=0$ for different magnetic fields $B$. We used the same physical parameters as in Fig.\ 3(c) of the main text,
but 
for more 
values of the magnetic field $B$.}
\label{fig: splitting field}
\end{figure}

With increasing magnetic field, the zero-bias peak in the
  conductance of the SIAM splits into two subpeaks, the position of
  which is approximately given by $V\approx\pm B$. It has long been of
  interest to have a quantitatively reliable value for the ``splitting
  field'' at which the peak splitting first becomes noticable.  The
  splitting field can be defined in two ways: (i) as the field $B_*$
  at which the number of local maxima changes from one to larger than
  one; or (ii) as the field $B_{\ast \ast}$ at which the maximum at
  zero bias turns into a minimum. In principle, these two fields need
  not coincide: if two side peaks emerge in the flanks of the
  zero-bias peak before the central maximum has turned into a minimum,
  $B_\ast$ would be smaller than $B_{\ast\ast}$. However, we would
  like to argue this does not occur in the present case, for which the
  mechanism for the peak splitting is well understood. The zero-bias
  conductance peak is computed as the sum of two peaks, one for spin
  up and one for spin down. These are pushed apart with increasing
  field. Once their spacing becomes comparable to their widths, their
  sum changes from showing a single to a double maximum, with a local
  minimum in between. This implies $B_\ast = B_{\ast \ast}$. Note,
  though, that for fields just above $B_{\ast \ast}$, the local
  minimum between the two maxima will still be extremely weak and the
  curve will look essentially flat there. The two maxima will become
  discernable as unambiguous ``peaks'' only at fields somewhat larger
  than $B_{\ast \ast}$. Therefore, if one attempts to estimate
  $B_\ast$ from (noisy) numerical data, by determining the field, say
  $B_\ast^{\rm sp}$, at which side peaks (sp) first become clearly
  noticable, this will always yield values somewhat larger than
  $B_\ast = B_{\ast \ast}$.

\Fig{fig: splitting field} shows our numerical results for the
  zero-temperature conductance as a function of voltage for different
  magnetic fields around $B\approx T_K$, for $U/\Gamma=12$, as in
  Fig.~3(c-d) of the main text. While the  curve for $B/T_K = 1$ exhibits
  a clear peak for non-zero voltage, this is not the case for
  $B/T_K=0.8 T_K$, and  the curve for $B/T_K= 0.9$ is a bit too noisy
  to unambigously identify a side peak. 
  We may therefore regard $B^{\rm sp}_* = T_K$ as a conservative upper 
  bound for the actual splitting field.  On the other hand, it is
  not possible to estimate
 $B_{\ast\ast}$ from our data.
  $B_{\ast \ast}$ is the field at which
  $- C_V =  \left[\frac{\partial^2}{\partial
      V^2}g(V)\right]_{V=0}$,
  the curvature of the conductance at zero bias, changes from negative
  to positive. However, extracting this curvature reliably from our
  data would require a level of numerical noise on the order 
of $0.1\%$, all   the more when tuning $B$ such that $C_V$ tends to zero.

Very recently, exact results for $C_V$ and hence $B_{\ast \ast}$
  have become available.  Filippone, Moca, von Delft and Mora (FMDM)
  \cite{Filippone17arXiv} have pointed out that $C_V$ can be extracted
  from the magnetic field dependence of the local spin and charge
  susceptibilities of the SIAM, which can be computed using the Bethe
  Ansatz. However, the formula which FMDM obtained for $C_V$ was
  incorrect due to a sign error in their calculations.  A correct
  formula for $C_V$ was first published by Oguri and Hewson
  \cite{Oguri17,*Oguri17_2,*Oguri17_3}, who showed that the
  Fermi-liquid relations discussed by FMDM could also be derived using
  Ward identities and the analytic and antisymmetry properties of the
  vertex function of the SIAM. Very recently FMDM reported (see
  version~2 of \cite{Filippone17arXiv}) that upon eliminating their
  sign mistake, their corrected formula for $C_V$ coincides with that
  of Oguri and Hewson. Moreover, NRG results by A. Weichselbaum,
 included in Appendix D of version~3 of \cite{Filippone17arXiv}, agree with
the corrected FL predictions for $C_V$. Incidentally, Figs.~8(c,d) 
of that analysis illustrates why extracting  $C_V$ 
from $g^{T=0}(V)$ would require an accuracy of order  
0.1\% for the numerical determination of the conductance
as function $V$.

In the Kondo limit $U/\Gamma \gg 1$, FMDM obtained a splitting
  field of $B_{\ast \ast} =0.75073 T_K^{(\chi)}$, where
  $T_K^{(\chi)}=\frac{1}{4\chi_s}$ is the Kondo scale defined via the
  zero-field, zero-temperature spin susceptibility.  As stated in the
  caption of Fig.~3 of the main text, $T_K^{(\chi)}$ is related to the
  Kondo temperature used in this work, defined via
  \mbox{$g(T\!=\!T_K,V\!=\!0)=\frac{1}{2}$}, by
  $T_K^{(\chi)}=T_K/1.04$ for the parameters used in Figs.~3 and
  \ref{fig: splitting field}.  (For a detailed discussion of various
  different definitions of $T_K$, see Ref.~\cite{Hanl14}.) Thus, the
  Fermi-liquid prediction for the splitting field translates to
  $B_{\ast \ast} = 0.72 T_K$. The fact that our upper bound estimate,
  $B_\ast^{\rm sp} = T_K$, is somewhat but not much larger than
  this value implies that our results are compatible with the slitting
  field predictions from Fermi liquid theory.

\end{document}